\documentclass[aps,prx,twocolumn,superscriptaddress,showpacs]{revtex4-2}

\usepackage{amsmath,amssymb,amsfonts}
\usepackage{graphicx}
\usepackage{bm}
\usepackage{braket}
\usepackage{booktabs}
\usepackage{hyperref}
\usepackage{physics}
\usepackage{xcolor}
\usepackage{kotex}
\usepackage{ulem}

\newcommand{\up}{\uparrow}
\newcommand{\dn}{\downarrow}
\newcommand{\bk}{\bm{k}}

\newcommand{\xik}{\xi_{\bk}}
\newcommand{\Jk}{J_{\bk}}
\newcommand{\sgn}{\mathrm{sgn}}

\begin{document}

\title{Topological Ising superconductivity in two-dimensional $p$-wave magnet}

\author{Kyoung-Min Kim}
\address{Asia Pacific Center for Theoretical Physics, Pohang 37673, Republic of Korea}
\affiliation{Department of Physics, Pohang University of Science and Technology, Pohang, Gyeongbuk 37673, Korea}
\author{GiBaik Sim}
\affiliation{School of Physics, The University of Melbourne, Parkville, VIC 3010, Australia}

\author{Moon Jip Park}
\email{moonjippark@hanyang.ac.kr}
\address{Department of Physics, Hanyang University, Seoul 04763, Republic of Korea}
\address{Research Institute for Natural Science and High Pressure, Hanyang University, Seoul, 04763, Republic of Korea}

\date{\today}

\begin{abstract}
Fermi-surface spin splitting generated by non-relativistic exchange fields provides a new route to topological superconductivity without relying on strong spin-orbit coupling. Here, we study superconducting instabilities of a square-lattice $p$-wave magnet with onsite and nearest-neighbour attractive interactions. The odd-parity exchange field removes inversion symmetry in the spin-split electronic structure, mixing singlet and triplet order parameters within a single symmetry channel. The leading instability is a mixed-parity A$_1$ Ising state, in which singlet components coexist with the $p_x$-wave triplet component, whose
$\bm d$-vector is locked along the exchange-field axis. As the nearest-neighbour attraction
grows, this Ising state undergoes a transition into a nodal topological superconducting
phase with Majorana edge modes protected by momentum-resolved winding numbers. These modes extend over finite momentum intervals bounded by the surface projections of bulk point nodes. We further show that a Zeeman field perpendicular to the exchange field can induce a $\mathbb{Z}_2$ topological superconducting phase. Our results identify $p$-wave magnets as a versatile testbed for topological superconductivity driven by non-relativistic spin splitting.
\end{abstract}

\maketitle

\begin{figure*}[t]
\centering
\includegraphics[width=1.4\columnwidth]{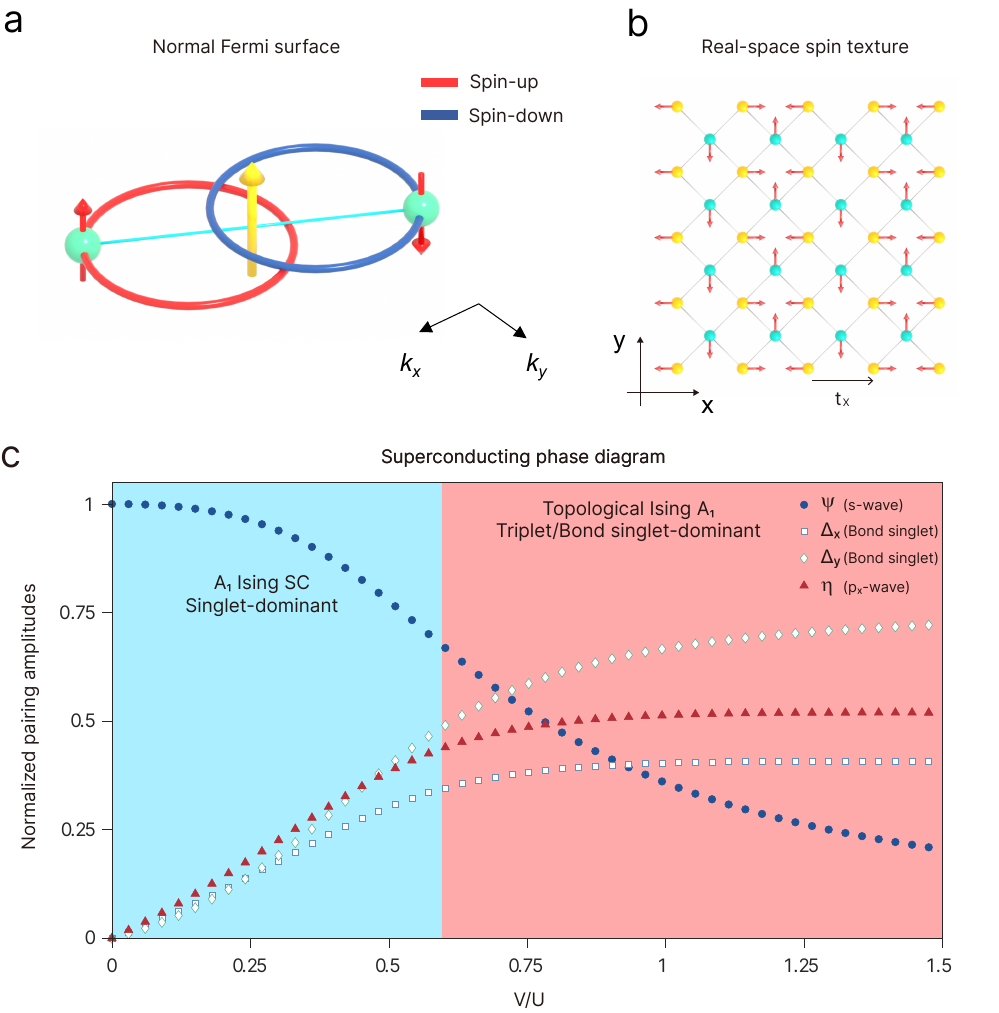}
\caption{\textbf{Ising superconductivity and phase diagrams in a $p$-wave magnet} \textbf{a}, Schematic representation of the pairing mechanism in a $p$-wave magnet. The red and blue contours denote the spin-up and spin-down Fermi surfaces in the $k_x$-$k_y$ plane, respectively. The yellow arrow represents the $\bm{d}$-vector, illustrating the orientation of the spin-triplet superconducting order parameter relative to the magnetic structure. \textbf{b}, Real-space lattice model of the $p$-wave magnet. Yellow and teal spheres represent the two sublattices $A$ and $B$. Red arrows indicate the local moments. $t_{\hat{x}}$ represents the half lattice translation vector \textbf{c}, Superconducting phase diagram as a function of $V/U$. The curves show the leading $A_1$ instability obtained from the linearized gap equations: the on-site $s$-wave $\psi$, the bond singlets $\Delta_x$ and $\Delta_y$, and the $p_x$-wave triplet $\eta$. The red region indicates the topological phase determined by non-trivial winding number. We used the parameter sets of $\mu=-1$, $J=2.5$, and $U=1$.}
\label{fig:1}
\end{figure*}

\section{Introduction.}

The search for topological superconductors has been a long-standing goal in condensed matter physics, driven by the quest to realize topologically protected Majorana modes \cite{RevModPhys.83.1057, Sato2017}. These boundary modes are intrinsically robust against local perturbations and hold promise for quantum information applications. Traditionally, the realization of topological superconductors has relied on three primary strategies: utilizing strong spin-orbit coupling (SOC) \cite{Oreg2010, Albrecht2016}, engineering proximity-induced heterostructures \cite{PhysRevLett.100.096407, doi:10.1126/science.1259327}, or discovering materials with intrinsic unconventional pairing symmetries \cite{RevModPhys.75.657}. However, these approaches often face fundamental challenges that hinder their practical implementation. For instance, the topological gap in SOC-based systems is fundamentally limited by the strength of the SOC, which is typically too small to provide the robust protection against the non-ideality in practical situations \cite{RevModPhys.87.137}. The limitation points to the need for an intrinsic platform where topology can be controlled by non-relativistic exchange splitting rather than weak relativistic SOC.

A promising direction to overcome these obstacles is provided by altermagnets~\cite{Smejkal2022Beyond, PhysRevX.12.040501}. Their nonrelativistic spin splitting can be large and originate from diverse microscopic mechanisms, including a $d$-wave Pomeranchuk instability, multipole order, and orbital ordering.~\cite{ahn2019antiferromagnetism,hayami2020bottom,yuan2020giant,jvv7-z2fq,leeb2024spontaneous,kaushal2025altermagnetism,durrnagel2025altermagnetic,Krempasky2024MnTe, Reimers2024CrSb} Along these lines, substantial effort has been devoted to realizing unconventional superconducting states in altermagnets, where the spin splitting can suppress conventional $s$-wave spin-singlet pairing.~\cite{ghorashi2024altermagnetic,PhysRevLett.134.146001,b7rh-v7nq,soto2014pair,zylh-rqxl,PhysRevB.110.L060508,52wh-1z5y,rhmg-j1fv,PhysRevB.110.205120,hu2025unconventional,liu2026fulde,llrq-1k9k,PhysRevB.108.205410,PhysRevB.109.224502,PhysRevLett.133.226002} One prominent direction has been the exploration of topological $p$-wave superconductivity~\cite{PhysRevB.108.184505,PhysRevB.108.224421,Hong2025,cadez2026emergencechiralpwavedwave,liu2025altermagnetism,fz2b-5sp7,Fukaya_2025}.

More recently, $p$-wave magnets have been proposed in a variety of settings, including orbital current states, localized spin models, Floquet engineered systems, van der Waals heterostructures, and other engineered platforms~\cite{Hellenes2024,Chakraborty2025,PhysRevLett.133.236703,zk69-k6b2,kim2026odd,sim2026quantum,ydjj-6r9l,c8pd-2fs4,li2026p,jin2026,xzm1-l6yf,9346-9jpf,7ywb-ml2q,Fukaya_2026}. Their odd-parity spin splitting breaks inversion symmetry, allowing spin-singlet and spin-triplet pairing to mix~\cite{GorkovRashba2001,Yip2014NCS}, and has been proposed as a route to topological superconductivity~\cite{PhysRevB.110.165429,cm1l-1rsh,PhysRevB.111.174519,pal2026emergentsuperconductingphasesunconventional,luo2026hiddenzeemanfieldoddparity}. In the large spin splitting limit, this mixing produces an Ising superconducting state with equal $s$-wave spin-singlet and $p$-wave spin-triplet components~\cite{Khodas2026}, in analogy with the Ising superconductivity established in transition-metal dichalcogenides~\cite{Yuan2014MoS2, Lu2015IsingMoS2,
Xi2016IsingNbSe2, Saito2016SpinValley, Zhou2016IsingMajorana}. However, a systematic study of this mixed $s$- and $p$-wave pairing in a microscopic lattice model beyond the large splitting limit, together with the topology of the resulting Ising superconducting state, has so far been missing.

In this manuscript, we show that a $p$-wave magnet can host topological superconductivity
through coupled singlet and triplet pairing. We analyze an extended
Hubbard model with attractive onsite ($U$) and nearest-neighbor ($V$) interactions. We find that the odd-parity exchange field locks
the singlet ($\psi$) and triplet ($\eta$) pairing amplitudes without enforcing an equal
mixture in the moderate-splitting regime. Instead, their relative weight evolves
continuously with $V/U$, interpolating between the singlet-dominated regime and the
strongly mixed Ising state, in contrast to the equal-mixing structure found in the strong
limit. The nearest-neighbour attraction additionally generates bond-singlet components
within the same $A_1$ channel, which play an essential role in the topological transition.
As $V/U$ grows, the resulting superconducting state enters a topological superconducting
phase characterized by the 1D DIII winding number. We further show that an external Zeeman
field perpendicular to the exchange axis induces the topological superconductivity even
when the singlet pairing dominates.

\section{Results}
\textbf{Minimal model of $p$-wave magnet.}--- 
The minimal two-band model of the itinerant electrons for the $p$-wave magnet is described by 
\begin{equation}
    h_0(\mathbf{k}) = \xi_{\bm{k}} +J_{\bm{k}} \sigma_z 
    \label{Eq:h0}
\end{equation}
where $\xi_{\bm{k}} = -2t(\cos k_x + \cos k_y) - \mu$ represents the kinetic dispersion. $J_{\bm{k}} = J\sin k_x$ denotes the $p$-wave exchange field. $\sigma_i$ is i-th Pauli matrix acting on spin space. We have chosen the exchange field axis along $z$-axis without loss of generality.  Throughout this work, we set the hopping parameter $t=1$ as the unit of energy. This exchange field $J_{\bm{k}}$ lifts the spin degeneracy, splitting the bands into $\epsilon_{\bm{k}\uparrow} = \xi_{\bm{k}} + J_{\bm{k}}$ and $\epsilon_{\bm{k}\downarrow} = \xi_{\bm{k}} - J_{\bm{k}}$ (Fig.~\ref{fig:1}\textbf{a}). Notably, the odd parity of $J_{\bm{k}}$ enforces the condition $\epsilon_{\bm{k}\uparrow} = \epsilon_{-\bm{k}\downarrow}$, which is a defining hallmark of $p$-wave magnetism~\cite{Hellenes2024}.

Microscopically, the effective model in Eq.~(1) can be derived from the parent lattice model in Fig.~\ref{fig:1}\textbf{b}, where itinerant electrons hop between two square sublattices with coplanar local moments rotating by $90^{\circ}$ per nearest-neighbour step. Projecting onto the low-energy bands yields the $p$-wave exchange field with $J \approx J_0^2/8t$ [Sec.~I of the SM~\cite{SM}] , where $J_0$ is the bare exchange coupling strength. In the absence of SOC, the symmetries of the magnetic state form a spin space group~\cite{Brinkman1966, Litvin1974, Liu2022SpinGroup, Smejkal2022Beyond}, the direct product of the spin-only group $\bm{r}_s = \{E, [\bar{C}_{2z}\|E]\}$ of the coplanar order and a nontrivial spin space group generated by the symmetries, $[C_{4z}\|t_{\bm\delta}]$, $[E\|\sigma_v(xz)]$, $[C_{2y_s}\|\sigma'_v(yz)]$. Here, $\bar C_{2z}$ denotes the combination of the twofold rotation $C_{2z}$ and spin inversion, $t_{\bm\delta}$ is the primitive lattice translation by the nearest-neighbour vector $\bm\delta$, and $y_s$ denotes the spin-space axis chosen along the direction of the local moments on sublattice $B$. Although the pure time reversal $[\mathcal{T}\|E]$ and inversion $[E\|P]$ are broken, the combined antiunitary symmetry $\mathcal{T}' = [\mathcal{T}\|t_{\hat{x}}]$ survives, whose projection onto the low-energy bands yields the effective time reversal $\Theta_{\mathrm{eff}} = i\sigma_y\mathcal{K}$ [Sec.~I of the SM~\cite{SM}]. The lattice parts of the residual elements form $C_{2v} = \{E, C_{2z}, \sigma_v(xz), \sigma'_v(yz)\}$, which governs the irreducible-representation classification of the pairing channels below.

\textbf{Symmetry classifications of superconducting channel.}---  To study the superconducting instabilities of the $p$-wave magnet, we take the effective two-band model in Eq. \eqref{Eq:h0} as the normal state, and we introduce the extended Hubbard model as the pairing interaction,
\begin{equation}
    H_\text{I} = -U\sum_i n_{i\up}n_{i\dn} - V\sum_{\langle i,j\rangle,\sigma,\sigma'} n_{i\sigma}n_{j\sigma'},
\label{eq:HI}
\end{equation}
where $n_{i\sigma} = c_{i\sigma}^\dagger c_{i\sigma}$ is the number operator at site $i$ with spin $\sigma$, $U > 0$ is the on-site attraction, and $V > 0$ is the nearest-neighbour attraction. The corresponding BdG Hamiltonian is 
\begin{equation}
H_{\mathrm{BdG}}
=\sum_{\mathbf{k}}
\Psi^{\dagger}_{\mathbf{k}}
\begin{pmatrix}
h_{0}(\mathbf{k}) & \Delta(\mathbf{k})\\
\Delta^{\dagger}(\mathbf{k}) & -h_{0}^{*}(-\mathbf{k})
\end{pmatrix}
\Psi_{\mathbf{k}},
\end{equation}
in the Nambu basis $\Psi_{\mathbf{k}}=(c_{\mathbf{k}\uparrow},c_{\mathbf{k}\downarrow},c^{\dagger}_{-\mathbf{k}\uparrow},c^{\dagger}_{-\mathbf{k}\downarrow})^{T}$. We write the superconducting gap matrix as
\begin{equation}
    \Delta(\bm{k}) = [\psi(\bm{k}) + \bm{d}(\bm{k}) \cdot \bm{\sigma}]\, i\sigma_y,
\end{equation}
where $\psi(\bm{k})$ and $\bm{d}(\bm{k})$ denote the spin-singlet and spin-triplet components, respectively.
Each component is expanded in the lattice basis functions $g_{\alpha}(\mathbf{k})$
classified by the irreducible representations (irreps) of the $C_{2v}$
point group [Table~\ref{tab:irreps}]. The symmetry-allowed channels are
\begin{equation}
\begin{array}{lll}
A_{1}: & \psi(\mathbf{k})=\psi\,g_{s}+\Delta_{x}\,g_{x}+\Delta_{y}\,g_{y}, &
\mathbf{d}(\mathbf{k})=\eta\,g_{p_{x}}\,\hat{z},\\[2pt]
B_{2}: & \psi(\mathbf{k})=0, &
\mathbf{d}(\mathbf{k})=\Delta_{p_{y}}\,g_{p_{y}}\,\hat{z},
\end{array}
\label{eq:irreps}
\end{equation}
where $\psi$ denotes the on-site $s$-wave singlet, $\Delta_x$ and $\Delta_y$ the
nearest-neighbor singlet amplitudes residing on the $x$ and $y$ bonds (bond singlets)~\cite{Micnas1990,Kotliar1988},
and $\eta$ and $\Delta_{p_y}$ the $p_x$- and $p_y$-wave spin-triplet amplitudes.

Notably, in the $A_{1}$ channel the broken inversion symmetry allows the
singlet ($\psi$, $\Delta_{x}$, $\Delta_{y}$) and triplet ($\eta$) components to mix
within a single gap matrix. We term this $A_{1}$ phase an ``Ising state''
in the sense that the spins of the constituent electrons and the
$\mathbf{d}$-vector of the Cooper pairs are locked along the direction of
the exchange field [Fig.~\ref{fig:1}\textbf{a}], analogous to conventional Ising
superconductors where these orientations are constrained to the
out-of-plane direction by atomic SOC~\cite{Lu2015IsingMoS2, Xi2016IsingNbSe2, Saito2016SpinValley,
Zhou2016IsingMajorana, Wickramaratne2023}. Note that in all
channels the $\mathbf{d}$-vector is oriented along $\hat{z}$. Other
orientations, such as $\mathbf{d}\parallel\hat{x}$ or $\hat{y}$,
correspond to equal-spin pairings and are energetically suppressed by the
exchange field, which favors the $z$-axis alignment as shown in
Fig.~\ref{fig:1}\textbf{a}.

Since all pairing channels considered in this study are of the inter-spin type
[$\bm{d}(\bm{k}) \parallel \hat{z}$], the $4\times 4$ BdG Hamiltonian
block-diagonalizes into two $2\times 2$ sectors spanned by
$\Psi_{+,\bm{k}} = (c_{\bm{k}\uparrow},
c^{\dagger}_{-\bm{k}\downarrow})^{T}$ and
$\Psi_{-,\bm{k}} = (c_{\bm{k}\downarrow},
-c^{\dagger}_{-\bm{k}\uparrow})^{T}$,
\begin{equation}
h^{\mathrm{BdG}}_{\lambda}(\bm{k}) =
\left(\xi_{\bm{k}} + \lambda J_{\bm{k}}\right)\tau_{z}
+ \Delta_{\lambda}(\bm{k})\,\tau_{x},
\label{eq:block}
\end{equation}
where $\lambda = \pm$ labels the two sectors,
$\tau_{i}$ are the Pauli matrices acting on the particle-hole space,
and $\Delta_{\lambda}(\bm{k}) = \psi(\bm{k}) + \lambda\, d_{z}(\bm{k})$
is the combined gap function of each sector. The quasiparticle energies
follow immediately as
\begin{equation}
E_{\lambda,\bm{k}} = \pm\sqrt{\left(\xi_{\bm{k}}
+ \lambda J_{\bm{k}}\right)^{2}
+ \left|\Delta_{\lambda}(\bm{k})\right|^{2}}.
\label{eq:E_bdg}
\end{equation}
Equation~\eqref{eq:block} makes the singlet-triplet mixing transparent at the
Hamiltonian level. Within the $A_1$ channel, each spin sector
experiences a single combined gap function
$\Delta_{\lambda}(\bm{k}) = \psi + \Delta_{x}\cos k_x + \Delta_{y}\cos k_y
+ \lambda\,\eta \sin k_x$.

\begin{table}[b]
\caption{
Pairing channels are classified according to the $C_{2v}$ irreducible representations
(irreps). The broken inversion symmetry merges singlet and triplet states into the same
$A_1$ irrep. The exchange field breaks the fourfold rotation, so the bond singlets
$\Delta_x$ and $\Delta_y$ belong to $A_1$ individually.}
\label{tab:irreps}
\begin{ruledtabular}
\begin{tabular}{lllll}
Irrep & $\Delta_\alpha$ & $g_\alpha(\bk)$ & Parity & Source \\
\hline
A$_1$ & $\psi$        & $g_s = 1$            & even & $U$ \\
A$_1$ & $\Delta_{x}$  & $g_{x} = \cos k_x$   & even & $V$ \\
A$_1$ & $\Delta_{y}$  & $g_{y} = \cos k_y$   & even & $V$ \\
A$_1$ & $\eta$        & $g_{p_x} = \sin k_x$ & odd  & $V$ \\
\hline
B$_2$ & $\Delta_{p_y}$ & $g_{p_y}=\sin k_y$  & odd  & $V$ \\
\end{tabular}
\end{ruledtabular}
\end{table}

\begin{figure*}[t]
\centering
\includegraphics[width=2\columnwidth]{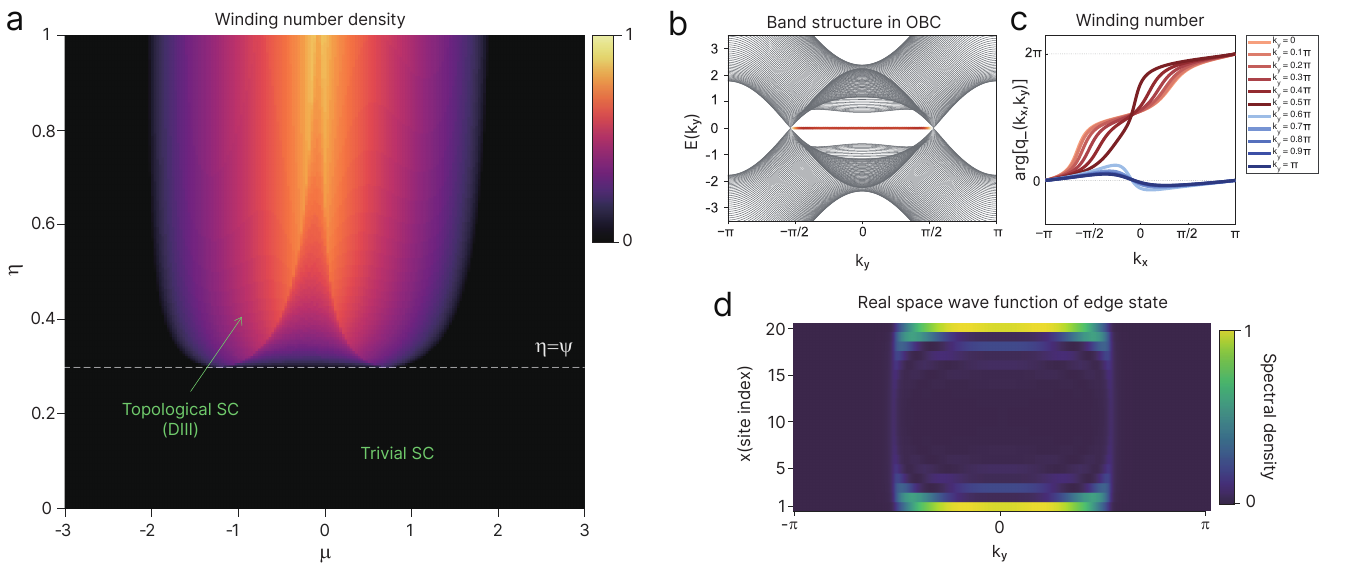}
\caption{
\textbf{Topological phase transition and bulk-boundary correspondence}. \textbf{a}, Topological phase diagram as a function of the chemical potential $\mu$ and the triplet pairing parameter $\eta$. The color scale represents the winding number density, clearly distinguishing the DIII-class topological superconducting (SC) phase (orange/yellow) from the trivial SC phase (black). The dashed teal line indicates the phase boundary where the bulk gap closes. \textbf{b}, Energy spectrum of the system in a slab geometry with open boundary conditions along the $x$-direction, plotted as a function of momentum $k_y$. Red dots highlight the zero-energy edge states emerging within the superconducting gap. \textbf{c},~Phase evolution of $q_{-}(k_x, k_y)$ in Eq.~\eqref{eq:winding} along
fixed-$k_y$ cuts. For cuts inside the topological window (red curves), the phase winds by $2\pi$ over the one-dimensional Brillouin zone, giving $|\nu(k_y)| = 1$, while cuts outside the window (blue curves) return to zero, $\nu(k_y) = 0$. \textbf{d}, Zero-energy spectral density $A(x, k_y, \omega=0)$ in the $(x, k_y)$ plane. The high-intensity peaks at the boundaries ($x = 1$ and $x = 20$) demonstrate the real-space localization of the topological edge modes, confirming the bulk-boundary correspondence. We used the parameters $J=1.0$, $\mu=-2.0$,
$\psi = 0.3$,
$\eta = 1.0$ for \textbf{b},\textbf{c},\textbf{d}.}

\label{fig:3}
\end{figure*}

\begin{figure*}[t]
\centering
\includegraphics[width=2\columnwidth]{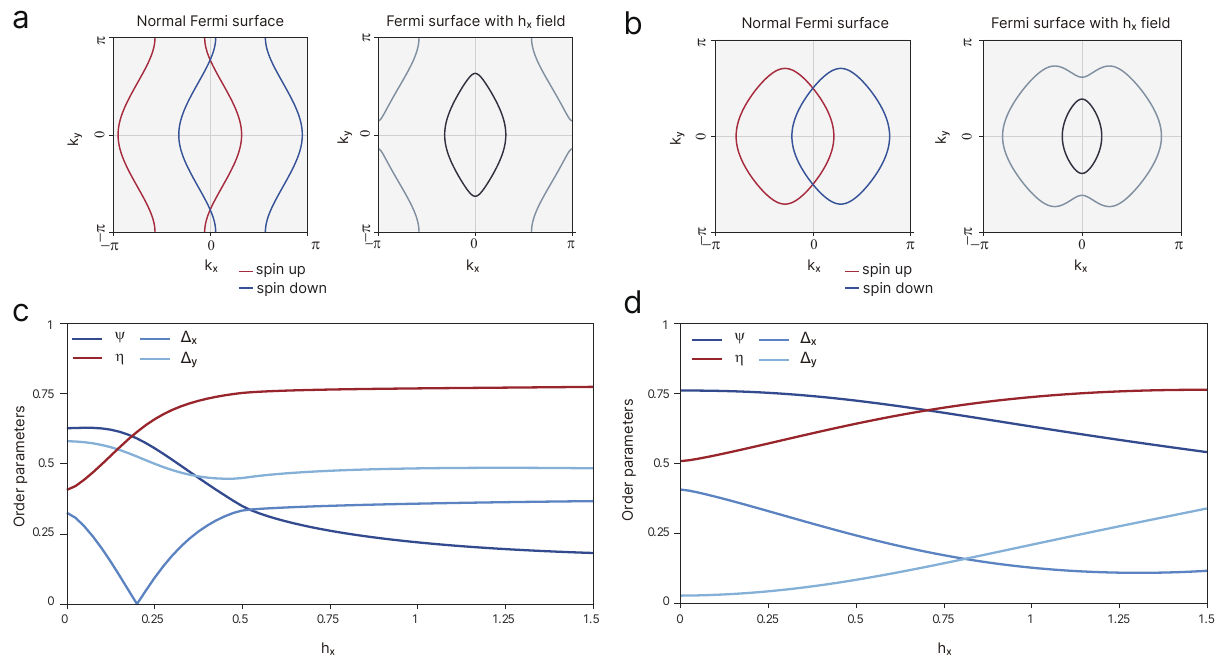}
\caption{\textbf{Field evolution of the mixed-parity Ising superconducting state}. \textbf{a, b} Spin-split Fermi surfaces of the $p$-wave magnet before and after applying a transverse Zeeman field $h_x$ for different chemical potentials $\mu=-0.5$, $\mu=-2$ respectively. Without $h_x$, the odd-parity exchange field shifts the two spin sectors in opposite directions in momentum space. A finite $h_x$ mixes the spin sectors and reconstructs the Fermi surface through avoided crossings. \textbf{c}, \textbf{d} Corresponding superconducting order parameters for the Fermi surface in \textbf{a,b}. The
on-site singlet $\psi$, the bond singlets $\Delta_x$, $\Delta_y$, and the $p_x$-wave
triplet $\eta$ remain in the same $A_1$ channel. The transverse field continuously
suppresses the on-site singlet and enhances the triplet. The parameters used are $J = 2.5$, $U = 1$, $V = 0.7$.
}
\label{fig:2}
\end{figure*}

\begin{figure*}[t]
\centering
\includegraphics[width=1.9\columnwidth]{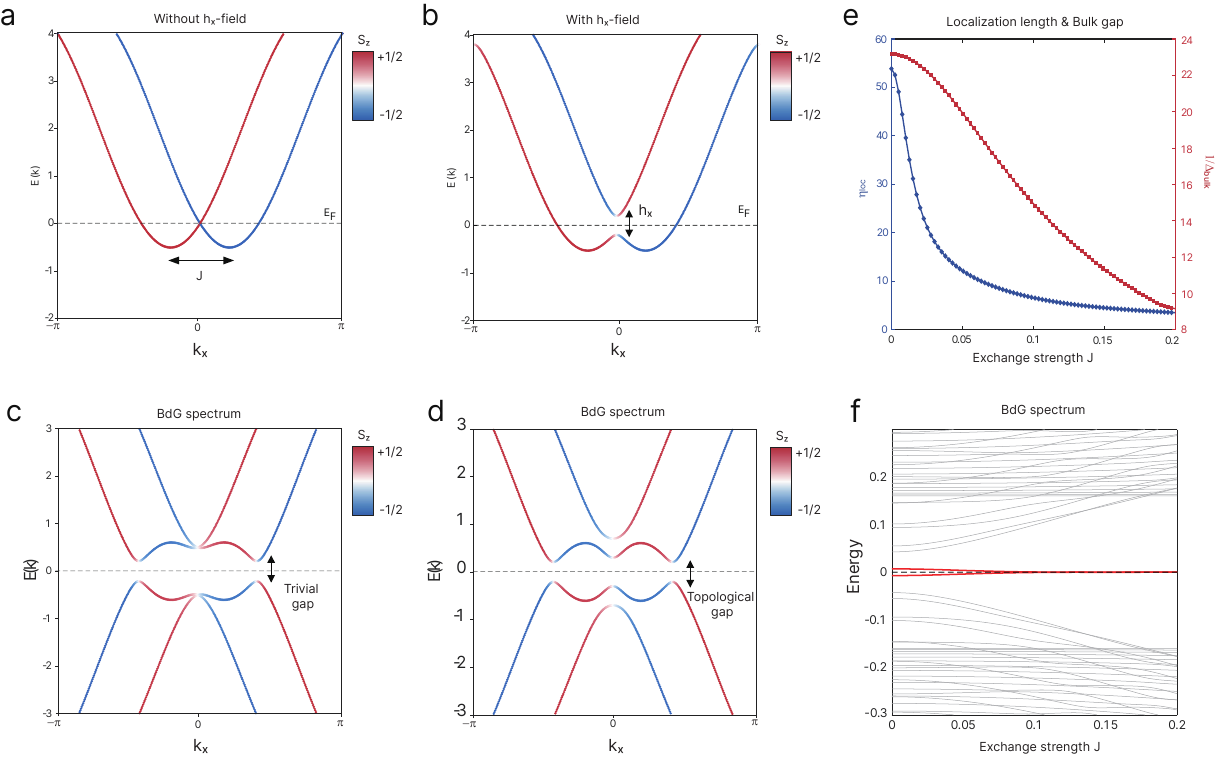}
\caption{
\textbf{Majorana states and localization with in-plane magnetic field.} 
\textbf{a},~\textbf{b},~Normal dispersions $E(k_x)$ without
(\textbf{a}) and with (\textbf{b}) the transverse field $h_x$.
$h_x$ shifts the two spin bands in
opposite directions along $k_x$. $h_x$ hybridizes them and opens a gap
at $k_x = 0$. The dashed line marks the Fermi level.
\textbf{c},~\textbf{d},~Corresponding BdG spectra. The pairing gap is
trivial at $h_x = 0$ (\textbf{c}) and topological at finite $h_x$
(\textbf{d}).
\textbf{e},~Localization length $\eta_{\mathrm{loc}}$ of the zero-energy
edge mode (left axis) and inverse bulk gap $1/\Delta_{\mathrm{bulk}}$
(right axis) as functions of the exchange strength $J$.
\textbf{f},~Finite-size spectrum versus $J$; red lines denote the
zero-energy Majorana edge states within the bulk gap (gray), confirming the existence of topologically protected Majorana modes. The parameters used are $J=1.5$, $\mu=-4$, $\psi=0.5$, $\eta=0.3$ with fixed $k_y=0$.
}
\label{fig:4}
\end{figure*}

\textbf{Superconducting phase diagram.}--- We solve the self-consistent linearized gap equations (Eqs.~\eqref{eq:gapeq1} and \eqref{eq:gapeq2} in the methods) near the critical temperature. The resulting order parameters determine both the dominant superconducting channel and the relative weights of the symmetry-allowed gap components. In particular, this approach captures the singlet-triplet mixing ratio $|\eta/\psi|$ of the $A_1$ state. 

Fig.~\ref{fig:1}\textbf{c} presents the resulting phase diagram as a function of $U$ and $V$. The $A_1$ Ising state is the leading instability over the entire range of $V/U$, while the $p_y$-wave B$_2$ channel is suppressed. At small $V/U$, the on-site $s$-wave $\psi$ is
dominant. The bond singlets $\Delta_x$, $\Delta_y$ and the triplet $\eta$ grow with $V$.
This structure is a direct consequence of the odd-parity exchange field $J$. Without $J$,
the extended Hubbard interaction couples $\psi$ only to the symmetric bond combination
$\Delta_x=\Delta_y$, while the $p_x$-wave channel remains in a separate odd-parity sector,
precluding it from entering the $A_1$ irrep. At finite $J$, the broken parity locks $\eta$
to the singlets, and the broken fourfold rotation additionally lifts the degeneracy between
the two bonds, $\Delta_x \neq \Delta_y$.

As $V/U$ increases, the weight is continuously transferred from the on-site singlet to the
bond singlets and the triplet. Interestingly, the topological transition occurs at $V/U \approx 0.6$, signified by the appearance of the bulk nodal point (Fig.\ref{fig:3}\textbf{b}),
where the on-site singlet $\psi$ is still the largest pairing component, and triplet $p$-wave is subdominant. As we show below, the bond singlets also play
an essential role in driving this transition.

\textbf{DIII Topological superconductivity.}--- In the topological superconducting phase, the $A_1$ state develops isolated point nodes in
the BdG spectrum. These nodes originate from the sign change of the pairing amplitude on
the Fermi surface. Along each fixed-$k_y$ line in the Brillouin zone, the gap function
$\Delta_{\lambda}(\bm{k}) = \psi + \Delta_x\cos k_x + \Delta_y\cos k_y
+ \lambda\,\eta\sin k_x$ changes sign as a function of $k_x$. $\Delta_y$ has the sign
opposite to $\psi$ and reduces the $k_x$-independent part of the gap, while $\Delta_x$ and
$\eta$ constitute the $k_x$-dependent oscillation. The sign change therefore occurs before
the triplet amplitude exceeds the on-site singlet. When the gap zeros intersect the Fermi
surface satisfying $\epsilon_{\bm{k}\pm}=\xi_{\bm{k}} \pm J_{\bm{k}} = 0$, the
quasiparticle energy in Eq.~\eqref{eq:E_bdg} vanishes, producing the point nodes in the
two-dimensional spectrum. These nodes exhibit a density of states that vanishes linearly,
$\rho(E) \propto |E|$, while the bulk gap remains finite at other $k_y$ values where the
Fermi surface does not intersect the gap zeros. In the limit of vanishing bond singlets,
$\Delta_x,\Delta_y\ll\psi,\eta$, the gap function reduces to
$\Delta(\bm{k}) = \psi + \eta\sin k_x$, which changes sign only in the triplet-dominant
regime $|\eta|>|\psi|$, with gap zeros at $\sin k_x^* = -\frac{\psi}{\eta}$,
$\cos k_y^* = -\frac{\mu + J\psi/\eta + 2t\sqrt{1 - (\psi/\eta)^2}}{2t}$~\cite{SatoFujimoto2009}.

The nodal structure motivates a $k_{y}$-resolved topological diagnosis~\cite{Sato2011FlatABS, Yada2011SurfaceDOS,
Schnyder2011FlatBands, Wong2013MajoranaFlat, Yuan2014MoS2,
Schnyder2015Review}. We can consider the spin-group mirror symmetry
$M_{y}=[E\|\sigma_v(xz)]$, which acts trivially on the spin degree of the freedom. Each
fixed-$k_{y}$ line therefore admits the composite time-reversal and particle-hole symmetries
$\Theta_{1\mathrm{D}}=M_{y}\Theta_{\mathrm{eff}}=i\sigma_y\tau_0 \mathcal{K}$ and
$\Xi_{1\mathrm{D}}=M_{y}\Xi=\sigma_0\tau_x \mathcal{K}$, satisfying
$\Theta_{1\mathrm{D}}^{2}=-1$ and $\Xi_{1\mathrm{D}}^{2}=+1$, where $\Xi$ is the particle-hole symmetry. $\mathcal{K}$ denotes complex conjugation The two symmetries place
every gapped cut in one-dimensional class DIII. Explicitly, the two operators act on the BdG Hamiltonian as,
\begin{align}
\Theta_{1{\rm D}} H_\textrm{BdG}(k_x,k_y)\Theta_{1{\rm D}}^{-1}
&= H_\textrm{BdG}(-k_x,k_y),
\\
\Xi_{1{\rm D}} H_\textrm{BdG}(k_x,k_y)\Xi_{1{\rm D}}^{-1}
&=- H_\textrm{BdG}(-k_x,k_y),
\nonumber
\end{align}
We compute the winding number of each BdG block
$h^{\mathrm{BdG}}_{\lambda}$ in Eq.~\eqref{eq:block},
\begin{equation}
\nu_{\lambda}(k_y) = \frac{1}{2\pi i}\oint dk_x\,
\partial_{k_x} \ln q_{\lambda}(k_x,k_y),
\label{eq:winding}
\end{equation}
where $q_{\lambda}(k_x,k_y) = (\xi_{\bm{k}} + \lambda J_{\bm{k}})
+ i\Delta_{\lambda}(\bm{k})$. The contour runs over the
one-dimensional Brillouin zone $k_x \in (-\pi,\pi]$. 
Due to the effective time-reversal symmetry $\Theta_{\mathrm{eff}}$, the two sectors carry opposite windings, $\nu(k_y) \equiv \nu_{-}(k_y)
= -\nu_{+}(k_y)$. This follows directly from the block structure of
Eq.~\eqref{eq:block}, $h^{\mathrm{BdG}}_{-}(k_x,k_y) =
h^{\mathrm{BdG}}_{+}(-k_x,k_y)$. $|\nu_{\lambda}|$
counts the number of protected zero modes per edge in each sector,
switching from $|\nu_{\lambda}| = 1$ to $0$ at the bulk point nodes
(Fig.~\ref{fig:3}\textbf{b},\textbf{c}). The total winding vanishes identically,
reflecting the absence of net chirality in class DIII, while each
sector contributes one zero mode per edge.
The two zero modes form a Majorana Kramers pair protected by
$\Theta_{\mathrm{1D}}^2 = -1$. Equivalently, the invariant of
each gapped $k_y$ cut is the parity $\nu \bmod 2$, consistent with the
$\mathbb{Z}_2$ classification of one-dimensional class DIII.

Integrating over the Brillouin zone, we define the winding number density as the fraction
of $k_y$ momenta carrying $|\nu| = 1$. To gain a heuristic understanding of the topological
phase, we evaluate the winding number for the two-component gap
$\Delta_{\lambda}(\bm{k}) = \psi + \lambda\,\eta\sin k_x$ with $\psi$ and $\eta$ treated as
free parameters. Fig.~\ref{fig:3}\textbf{a} shows the winding number density, defined as
$\sum_{k_y}|\nu(k_y)|/N_{k_y}$ across the $(\mu,\,\eta)$ plane, where $N_{k_y}$ is the number of sampled $k_y$ momenta. The topological
superconducting phase is sharply distinguished from the trivial phase by non-zero winding number density. To establish the bulk--boundary correspondence, we compute the quasiparticle spectrum in a slab geometry with open boundaries along $x$ (Fig.~\ref{fig:3}\textbf{b}). The zero-energy states appear within the superconducting gap, and their momentum support coincides precisely with the $k_y$ window where $\nu=1$ (see also Fig.~S2 in the SM~\cite{SM} for the corresponding edge-spectrum and winding-number analysis under open boundary conditions). The real-space localization of these modes is further confirmed by the zero-energy spectral density $A(x, k_y, \omega = 0)$ [Fig.~\ref{fig:3}\textbf{d}], which displays sharp intensity peaks concentrated at the sample boundaries $x = 1$ and $x = 20$ throughout the topological window of $k_y$, confirming the bulk-boundary correspondence. The phase diagram remains qualitatively unchanged even when the bond
singlet components $\Delta_x$ or $\Delta_y$ are included. The
notable effect of $\Delta_y$ is to push the topological phase boundary
toward smaller $\psi$, as it reduces the constant ($k_y$-independent)
part of the gap function discussed above, whereas $\Delta_x$ slightly
expands the topological phase.

\textbf{Effect of external magnetic field.}--- The singlet-triplet balance of the $A_1$ state can also be tuned externally by an in-plane Zeeman field $h_x$ perpendicular to the exchange axis~\cite{Frigeri2004,Smidman2017,Yip2014NCS}. In contrast to the momentum-dependent $p$-wave exchange field $J\sin k_x\sigma_z$, which preserves the spin sectors and merely shifts them in opposite directions along $k_x$ [Figs.~\ref{fig:2}\textbf{a} and \textbf{b}; left], the transverse Zeeman field $h_x\sigma_x$ mixes the two spin sectors. This mixing reconnects the displaced Fermi pockets through avoided crossings and opens a finite gap along the original spin-degeneracy line [Figs.~\ref{fig:2}\textbf{a} and \textbf{b}; right]. Such Fermi-surface reconstruction reshuffles the pairing channels
of the $A_1$ state. Figs.~\ref{fig:2}\textbf{c},\textbf{d} trace the
singlet $\psi$, the bond singlets $\Delta_x$, $\Delta_y$, and the triplet
$\eta$ as $h_x$ is increased, for two representative chemical
potentials $\mu=-0.5$ and $\mu=-2$. In both cases the onsite singlet $\psi$ is
continuously suppressed while the triplet grows. The four components remain
symmetry-locked throughout. $h_x$ redistributes weight inside the
$A_1$ irrep without splitting it. The transverse field thus provides
a complementary route to realize the triplet dominant phase. A
field-induced equal-spin component $\hat{y}\sin k_x$~\cite{Khodas2026}
vanishes at the time-reversal invariant momenta and does not affect the
topological criterion below. At the
same time, $h_x$ breaks the effective time-reversal symmetry
$\Theta_{\mathrm{eff}}$ and reduces the topological classification
from class DIII to class D.

Within this class-D regime, the topological invariant at each $k_y$ 
is determined by the Pfaffian as,
\begin{equation}
\begin{aligned}
    \mathcal{M}(k_y) = & \; \mathrm{sgn}\!\left[\mathrm{Pf}\!\big(\tau_x \tilde{H}_\textrm{BdG}(k_x\!=\!0, k_y)\big)\right]\\
   & \times \mathrm{sgn}\!\left[\mathrm{Pf}\!\big(\tau_x \tilde{H}_\textrm{BdG}(k_x\!=\!\pi, k_y)\big)\right].
\label{eq:majorana}
\end{aligned}
\end{equation}
where $\tau$ is the Pauli matrix acting on particle-hole basis.
$\tilde{H}_{\mathrm{BdG}}(\bm{k}) = \xi_{\bm k}\tau_z\sigma_0 + J\sin k_x\,\tau_0\sigma_z
+ h_x\,\tau_z\sigma_x - \psi(\bm{k})\,\tau_y\sigma_y + \eta\sin k_x\,\tau_x\sigma_x$ is the
BdG Hamiltonian with the singlet gap function $\psi(\bm{k})$ defined in
Eq.~\eqref{eq:irreps}. At $k_x^* = 0,\pi$, both $J\sin k_x$ and $\eta\sin k_x$ vanish, and
the singlet gap reduces to
$ \psi \pm \Delta_x + \Delta_y\cos k_y$. The Pfaffian then evaluates to
\begin{equation}
\mathrm{Pf}\big(\tau_x\tilde{H}_{\mathrm{BdG}}(k_x^*,k_y)\big)
= |h_x|^2 - \xi_{k_x^*}(k_y)^2 - \psi_{k_x^*}(k_y)^2,
\end{equation}
with $\xi_{k_x^*}(k_y) = -2t(\cos k_x^* + \cos k_y) - \mu$. The condition for non-trivial
$\mathcal{M} = -1$ then requires
\begin{equation}
\sqrt{\xi_0(k_y)^2 + \psi_0(k_y)^2} < |h_x| < \sqrt{\xi_\pi(k_y)^2 + \psi_\pi(k_y)^2}.
\end{equation}
This gap-closing condition is 
identical to the well-known criterion for topological Majorana modes 
in spin-orbit-coupled semiconductor 
nanowires~\cite{Lutchyn2010,Oreg2010}, with the exchange field 
$J\sin k_x$ playing the role of the Rashba SOC $\alpha_R k$ [Fig. \ref{fig:4}]. The 
key difference is the underlying energy scale: the Rashba SOC is 
relativistic in origin and typically much smaller than the bandwidth, 
whereas the exchange splitting $J$ in $p$-wave magnets can be  
non-relativistic, yielding a 
proportionally larger topological gap 
$\Delta_{\mathrm{topo}} \sim J\psi/h_x \gg \alpha_R k_F \Delta_s/V_Z$ as shown in Fig. \ref{fig:4}\textbf{f}.

The Pfaffian criterion fixes the topological invariant from the two
TRIM, but the bulk gap at generic $k_x$ is anisotropic. The exchange
field $J\sin k_x\,\sigma_z$ breaks SU(2), and the in-plane component
$h_x\sigma_x$ anticommutes with $J\sigma_z$, acting as an effective
spin-orbit coupling that protects the gap away from the high-symmetry
lines (see Fig.~S3 in the SM~\cite{SM} for the evolution of the slab
BdG spectrum with the effective spin-orbit-coupling strength $\alpha_R$). Figs.~\ref{fig:4}\textbf{a},\textbf{b} compare the normal-state
dispersions without and with $h_x$, and Figs.~\ref{fig:4}\textbf{c},
\textbf{d} the corresponding BdG spectra, illustrating the opening
of a topological gap.

Within the topological regime, the zero-energy modes are sharply 
localised at the system boundaries, and their localisation length 
decreases exponentially with $J$, tracking the inverse bulk gap 
$1/\Delta_{\mathrm{bulk}}$ (Fig.~\ref{fig:4}\textbf{e}). The 
finite-size spectrum (Fig.~\ref{fig:4}\textbf{f}) confirms that 
the zero modes persist robustly across the entire topological 
window of $J$.

\section{Discussions.}
We have shown that the Ising superconducting state of a $p$-wave 
magnet harbors a topological transition controlled by the 
interplay of the on-site singlet, the bond singlets, and the 
triplet within the single $A_1$ channel. The essential role of the 
$p$-wave exchange field is twofold. First, it locks the singlet 
and triplet into a single $A_1$ channel, so that the $s+p$ Ising state forms as a 
single stable phase for any $U, V > 0$. Second, by absorbing 
the exchange splitting into the quasiparticle dispersion rather 
than the pairing sector, it eliminates the Bogoliubov Fermi surface 
that would otherwise suppress the condensation energy. As a result, 
the $A_1$ channel is energetically preferred over equal-spin triplet 
states~\cite{Hong2025}. Together, these features allow the 
nearest-neighbour interaction $V$ to drive a topological transition 
within the singlet-dominant regime, without requiring a 
change in pairing symmetry or a separate triplet instability.

The field-induced topological phase places $p$-wave magnets in 
direct correspondence with the semiconductor nanowire route to 
Majorana modes~\cite{Lutchyn2010,Oreg2010}, with the 
non-relativistic exchange scale $J$ replacing the relativistic 
Rashba SOC, which may offer more robust topological protections. Our results 
suggest that the odd-parity exchange splitting characteristic 
of $p$-wave magnets functions as a 
promising ingredient for engineering topological superconductivity in 
magnetically ordered systems.

\section{Methods}

The pairing amplitudes $\psi$, $\Delta_x$, $\Delta_y$, and $\eta$ are determined by the
following coupled self-consistent gap equations:
\begin{equation}
\begin{pmatrix} \psi \\ \Delta_{x} \\ \Delta_{y} \\ \eta \end{pmatrix} =
\begin{pmatrix}
U\mathcal{I}_{s,s} & U\mathcal{I}_{s,x} & U\mathcal{I}_{s,y} & U\mathcal{I}_{s,p_x} \\
2V\mathcal{I}_{s,x} & 2V\mathcal{I}_{x,x} & 2V\mathcal{I}_{x,y} & 2V\mathcal{I}_{x,p_x} \\
2V\mathcal{I}_{s,y} & 2V\mathcal{I}_{x,y} & 2V\mathcal{I}_{y,y} & 2V\mathcal{I}_{y,p_x} \\
2V\mathcal{I}_{s,p_x} & 2V\mathcal{I}_{x,p_x} & 2V\mathcal{I}_{y,p_x} & 2V\mathcal{I}_{p_x,p_x}
\end{pmatrix}
\begin{pmatrix} \psi \\ \Delta_{x} \\ \Delta_{y} \\ \eta \end{pmatrix},
\label{eq:gapeq1}
\end{equation}
whereas the B$_2$ amplitude satisfies the decoupled equation
\begin{equation}
\Delta_{p_y} = 2V \mathcal{I}_{p_y,p_y} \Delta_{p_y}.
\label{eq:gapeq2}
\end{equation}
The overlap integrals are defined for $\alpha,\beta \in \{s, x, y, p_x, p_y\}$ as,
\begin{equation}
\mathcal{I}_{\alpha,\beta}(T) = \frac{1}{N}\sum_{\bm{k}}
\frac{g_\alpha(\bm{k})g_\beta(\bm{k})}{2E_{\bm{k}}}
\tanh\left(\frac{E_{\bm{k}}}{2T}\right).
\label{eq:Iab}
\end{equation}
Here $N$ is the number of lattice sites. $E_{\bm{k}}$ is the normal state energy. (See supplementary material for the detailed gap equation.)

Notably, the off-diagonal structure of the matrix in Eq.~\eqref{eq:gapeq1} explicitly
encodes the $s$-$p$ locking mechanism, which originates from the inversion symmetry
breaking inherent to $p$-wave magnets. At $J=0$, the quasiparticle spectrum is even under
$k_x\to -k_x$. Consequently, the integrands of the cross terms $\mathcal{I}_{s,p_x}$,
$\mathcal{I}_{x,p_x}$, and $\mathcal{I}_{y,p_x}$ are odd in $k_x$, because they contain the
odd-parity basis function $g_{p_x}(\bm{k})=\sin k_x$ while all other factors are even.
These terms vanish upon integration over the Brillouin zone, so the spin-singlet and
spin-triplet pairing sectors are symmetry-decoupled in this limit. The spectrum at $J=0$ is
also symmetric under $k_x\leftrightarrow k_y$, which enforces
$\mathcal{I}_{s,x}=\mathcal{I}_{s,y}$ and $\mathcal{I}_{x,x}=\mathcal{I}_{y,y}$, so that
the antisymmetric bond combination $\Delta_x-\Delta_y$ decouples from the $s$-wave sector.
Conversely, at finite $J$, the $p$-wave exchange field $J_{\bm{k}}$ introduces an
asymmetric shift of the Fermi surface along $k_x$. This breaks both cancellations, locking
the triplet to the singlets and inducing the bond anisotropy $\Delta_x\neq\Delta_y$. For a
detailed derivation of these gap equations, see Sec.~III of the SM~\cite{SM}.

The linearized gap equations in Eq.~\eqref{eq:gapeq1} simplify analytically upon projecting onto the dominant $\psi$--$\eta$ sector. 
Within this regime, $\Delta_{A_1}(\bk) = \psi + \eta\,\sin k_x$, and its corresponding gap equation becomes:
\begin{equation}
\begin{pmatrix} \psi \\ \eta \end{pmatrix}
= \begin{pmatrix}
U\tilde{\mathcal{I}}_{s} & U\tilde{\mathcal{I}}_{st} \\
2V\tilde{\mathcal{I}}_{st} & 2V\tilde{\mathcal{I}}_{t}
\end{pmatrix}
\begin{pmatrix} \psi \\ \eta \end{pmatrix},
\label{eq:gapeq1_reduced}
\end{equation}
where $\tilde{\mathcal{I}}_{s}=\mathcal{I}_{s,s}$, $\tilde{\mathcal{I}}_{st}=\mathcal{I}_{s,p_x}$, and $\tilde{\mathcal{I}}_{t}=\mathcal{I}_{p_x,p_x}$ are now independent of $\psi$ and $\eta$.
The reduced gap equation at $T_c$ yields the leading eigenvalue
\begin{equation}
    \lambda_+ = \tfrac{1}{2}(U\tilde{\mathcal{I}}_{s} + 2V\tilde{\mathcal{I}}_{t}) + \tfrac{1}{2}\sqrt{(U\tilde{\mathcal{I}}_{s} - 2V\tilde{\mathcal{I}}_{t})^2
  + 8UV\tilde{\mathcal{I}}_{st}^2},
\label{eq:lambda}
\end{equation}
where the integrals are evaluated at $\Delta(\bm{k}) = 0$ within linearized gap equation formalism. The cross-term $8UV\tilde{\mathcal{I}}_{st}^2$, present only at finite $J$, unconditionally increases $\lambda_+$. We find that the $s$-$p$ locking mechanism always enhances $T_c$, regardless of the ratio $V/U$.
The corresponding eigenvector gives the gap ratio
\begin{equation}
    \frac{\eta}{\psi} = \frac{2V\tilde{\mathcal{I}}_{st}}{\tfrac{1}{2}  (U\tilde{\mathcal{I}}_{s} - 2V\tilde{\mathcal{I}}_{t}) + \tfrac{1}{2}\sqrt{(U\tilde{\mathcal{I}}_{s} - 2V\tilde{\mathcal{I}}_{t})^2 + 8UV\tilde{\mathcal{I}}_{st}^2}},
\label{eq:ratio}
\end{equation}
which interpolates continuously from the singlet-dominant regime
($|\eta/\psi| \ll 1$ for $U \gg V$) to the triplet-dominant regime
($|\eta/\psi| \gg 1$ for $V \gg U$). The topological transition is reached at $|\eta/\psi| = 1$,
confirming that the topological transition lies within the
physically accessible region of the Ising phase.

\newpage
\begin{acknowledgments}
This work was supported by the National Research Foundation of Korea (NRF) grant funded by the Korea government (MSIT) (Grants No. RS-2025-16070482, RS-2025-25464760, RS-2026-25519864, RS-2025-25446099, RS-2023-NR119928, RS-2025-03392969, RS-2026-25607155). This work was also supported by BK21 FOUR (Fostering Outstanding Universities for Research) program through the National Research Foundation (NRF) funded by the Ministry of Education of Korea. K. K. was supported by an appointment to the JRG Program at the APCTP through the Science and Technology Promotion Fund and Lottery Fund of the Korean Government, the Korean Local Governments (Gyeongsangbuk-do Province and Pohang City), and the National Research Foundation of Korea (NRF) funded by the Korean government (Ministry of Science and ICT, MSIT) (No. RS-2026-25499525). G.B.S. was supported by the Australian Research Council (ARC) through Grant No.\ DP240100168 and the NRF through Grant No.\ RS-2024-00453943.

\end{acknowledgments}

\bibliography{bib_revised}

\newpage
\pagebreak


\onecolumngrid
\clearpage

\setcounter{equation}{0}
\setcounter{figure}{0}
\setcounter{table}{0}
\setcounter{section}{0}
\renewcommand{\theequation}{S\arabic{equation}}
\renewcommand{\thefigure}{S\arabic{figure}}
\renewcommand{\thetable}{S\arabic{table}}
\renewcommand{\thesection}{S\arabic{section}}
\renewcommand{\theHequation}{S\arabic{equation}}
\renewcommand{\theHfigure}{S\arabic{figure}}
\renewcommand{\theHtable}{S\arabic{table}}
\renewcommand{\theHsection}{S\arabic{section}}

\begin{center}
{\large \textbf{Supplemental Material for}}\\[0.3em]
{\large \textbf{``Topological Ising superconductivity in two-dimensional $p$-wave magnet''}}\\[0.5em]
{Kyoung-Min Kim, Gibaik Sim, and Moon Jip Park}
\end{center}

\vspace{1em}

\section{Microscopic lattice model}
\label{sec:4band}

\begin{figure*}[b]
\centering
\includegraphics[width=1\linewidth]{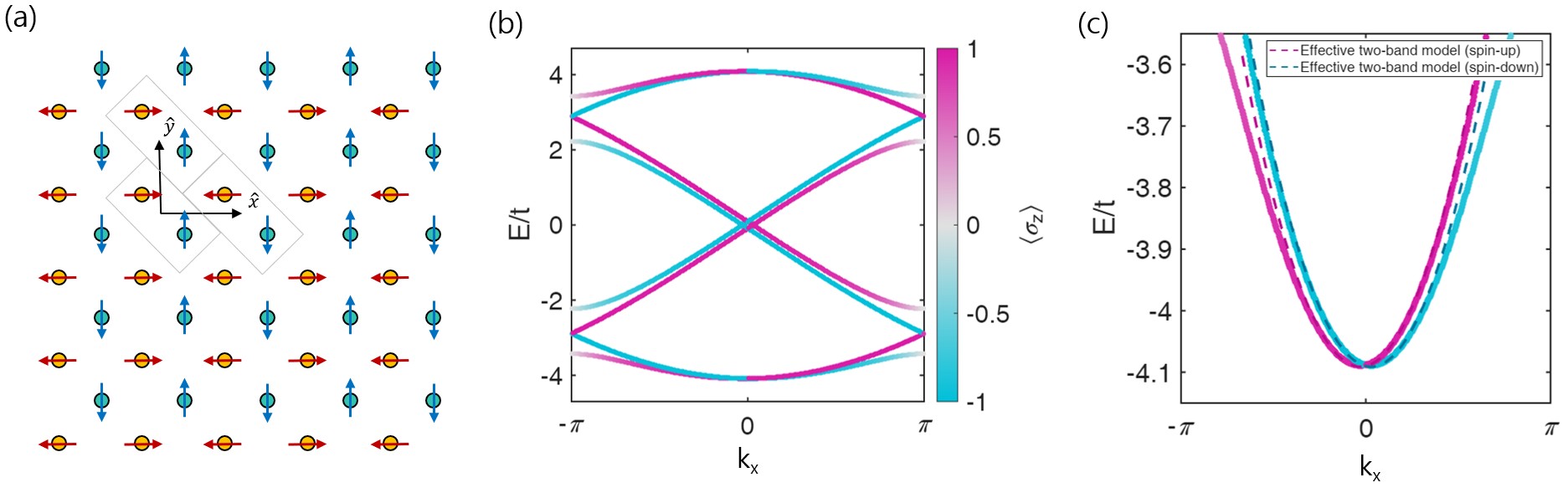}
\caption{
(a)~Schematic of the lattice model, Eq.~\eqref{eq:full_lattice}. Each sublattice
orders antiferromagnetically along $\hat{x}$ with mutually orthogonal ordering
axes.
(b)~Band structure of the magnetic-cell Hamiltonian at $k_{y}=0$. The bands split into
oppositely polarized branches with a splitting odd in $k_{x}$.
(c)~Magnified view near the band bottom, compared with the effective
two-band model (dashed). Parameters used: $t=1$, $J_{sd}=0.6$.}
\label{figs:lattice}
\end{figure*}

We consider the two-dimensional lattice composed of two magnetic sublattices $A$ (yellow) and $B$ (teal) as shown in Fig. \ref{figs:lattice}. The magnetic moments $\hat{\mathbf{m}}_A$ and $\hat{\mathbf{m}}_B$ at site $(i,j)$ are given by
\begin{equation}
\hat{\mathbf{m}}_{A, (i,j)} = (-1)^i \hat{x}, \quad \hat{\mathbf{m}}_{B, (i,j)} = (-1)^i \hat{y},
\end{equation}
where the moments on sublattice $A$ point along the horizontal direction ($\pm\hat{x}$), while those on sublattice $B$ point along the vertical direction ($\pm\hat{y}$). Both sublattices exhibit an antiferromagnetic ordering along the $\hat{x}$-direction indexed by $i$.

The tight-binding Hamiltonian with nearest-neighbor hopping between sublattices $A$ and $B$ can be written as,
\begin{equation}
H = H_0+V
\label{eq:full_lattice}
\end{equation}
\begin{gather}
H_{0}=
-t\sum_{i,j}\sum_{\sigma=\uparrow,\downarrow}
\Big[
 \big(c^{\dagger}_{i+1,j,A,\sigma}+c^{\dagger}_{i,j-1,A,\sigma}
     +c^{\dagger}_{i,j,A,\sigma}+c^{\dagger}_{i+1,j-1,A,\sigma}\big)
     \,c_{i,j,B,\sigma}\\+
 \big(c^{\dagger}_{i-1,j,B,\sigma}+c^{\dagger}_{i,j+1,B,\sigma}
     +c^{\dagger}_{i,j,B,\sigma}+c^{\dagger}_{i-1,j+1,B,\sigma}\big)
     \,c_{i,j,A,\sigma}
     \nonumber
\Big]
\end{gather}
\begin{equation}
V= J_{0} \sum_{\alpha=A,B}\sum_{i,j} c_{i,j,\alpha}^\dagger (\hat{\mathbf{m}}_{\alpha,(i,j)} \cdot \bm{\sigma}) c_{i,j,\alpha},
\end{equation}
where $c_{i,j,\alpha,\sigma}^\dagger$ ($c_{i,j,\alpha,\sigma}$) creates (annihilates) an electron with spin $\sigma \in \{\uparrow, \downarrow\}$ on sublattice $\alpha \in \{A, B\}$ at position $(i,j)$. $t$ is the nearest-neighbor hopping amplitude between $A$ and $B$ sites, and $J_0$ is the exchange coupling strength to local magnetic moments $\hat{\mathbf{m}}_{\alpha,(i,j)}$. Here, $c_{i,j,\alpha}^\dagger = (c_{i,j,\alpha,\uparrow}^\dagger, c_{i,j,\alpha,\downarrow}^\dagger)$ denotes the spinor operator. Although physical time-reversal symmetry $\mathcal{T} = i\sigma_y \mathcal{K}$ is broken by the magnetic order, where $\mathcal{K}$ is the complex conjugation, the model in Eq. \eqref{eq:full_lattice} possesses a nonsymmorphic rotation and time-reversal symmetry respectively,
\begin{equation}
C_{2}'=[C_{2z}||t_{\hat{x}} ], \quad \mathcal{T}' = \mathcal{T}  t_{\hat{x}},
\label{eq:Tprime}
\end{equation}
where $t_{\hat{x}}$ denotes a lattice translation along $\hat{x}$-axis. We note that the coplanar texture breaks the continuous spin-rotation symmetry about the $z$ axis, so that the $z$-directional spin is not a conserved quantity of the lattice model. Nevertheless, as we show below,
the two low-energy bands acquire a momentum-odd splitting whose spin
expectation value (not a quantum number) is polarized along
$\sigma_{z}$, which is the defining feature of a $p$-wave magnet.  More precisely, defining the spin-resolved spectral function
$A_{\sigma}(\mathbf{k},\omega)=\sum_{\alpha,n}|\langle n|
c^{\dagger}_{\mathbf{k}\alpha\sigma}|0\rangle|^{2}\delta(\omega-E_{n})$,
where $|n\rangle$ are the exact eigenstates and $|0\rangle$ the ground
state, the antiunitary $\mathcal{T}'$ enforces the identity
\begin{equation}
A_{\sigma}(\mathbf{k},\omega)=A_{\bar\sigma}(-\mathbf{k},\omega)
\end{equation}

Formally, the full spin space group of the magnetic state is
the direct product of the spin-only group and the nontrivial spin space
group~\cite{Brinkman1966, Litvin1974, Liu2022SpinGroup,
Smejkal2022Beyond}. The coplanar texture yields the spin-only group
$\bm{r}_s = \{E, [\bar{C}_{2z}\|E]\}$, where $\bar{C}_{2z}$ combines
$C_{2z}$ with time reversal. The nontrivial spin space
group is generated by the three elements
\begin{equation}
[C_{4z} \,\|\, t_{\bm{\delta}}], \qquad
[E \,\|\, \sigma_v(xz)], \qquad
[C_{2y_s} \,\|\, \sigma'_v(yz)],
\end{equation}
where $\bm{\delta} = (\hat{x}+\hat{y})/2$ is the nearest-neighbour
translation. The square of first generator gives $[C_{2z}\|t_{\hat{x}+\hat{y}}]$, which equals the
nonsymmorphic rotation $C_2' = [C_{2z}\|t_{\hat{x}}]$ of Eq.~(S5) up to
the magnetic lattice translation $t_{\hat{y}}$.
combining $C_2'$ with $\bar{C}_{2z} \in \bm{r}_s$ yields the effective
time reversal $\mathcal{T}' = [\mathcal{T}\|t_{\hat{x}}]$. The pure
operations $[\mathcal{T}\|E]$, $[E\|t_{\hat{x}}]$, $[E\|P]$, and the four-fold
rotation $[g_s\|C_{4z}]$ are broken for any spin operation $g_s$,
reducing the lattice point group from $C_{4v}$ to $C_{2v}$.

Although the rotating magnetic moment doubles the unit
cell, the periodicity can be restored by rotating the spin frame at
each site to its local moment,
$
c_{i,j,\alpha}
=e^{-i\theta_{\alpha,(i,j)}\sigma_{z}/2}
d_{i,j,\alpha},
$
where $\theta_{\alpha,(i,j)}$ is the in-plane angle of the local moment
$\hat{\mathbf{m}}_{\alpha,(i,j)}$. The transformed Hamiltonian reads
\begin{align}
H_{0}=-t\sum_{i,j}\sum_{\sigma=\uparrow,\downarrow}
\Big[&
e^{+i\sigma\pi/4}
\big(d^{\dagger}_{i,j,A,\sigma}+d^{\dagger}_{i,j-1,A,\sigma}\big)
d_{i,j,B,\sigma}
+e^{-i\sigma\pi/4}
\big(d^{\dagger}_{i+1,j,A,\sigma}+d^{\dagger}_{i+1,j-1,A,\sigma}\big)
d_{i,j,B,\sigma}
\nonumber\\
+&e^{-i\sigma\pi/4}
\big(d^{\dagger}_{i,j,B,\sigma}+d^{\dagger}_{i,j+1,B,\sigma}\big)
d_{i,j,A,\sigma}
+e^{+i\sigma\pi/4}
\big(d^{\dagger}_{i-1,j,B,\sigma}+d^{\dagger}_{i-1,j+1,B,\sigma}\big)
d_{i,j,A,\sigma}\Big],
\label{eq:Hd}
\\[2pt]
V=&\;J_{0}\sum_{\alpha=A,B}\sum_{i,j}
d^{\dagger}_{i,j,\alpha}\,\sigma_{x}\,d_{i,j,\alpha}.
\label{eq:Vd}
\end{align}

Fourier transformation of the above Hamiltonian gives
the four-band Bloch Hamiltonian,
\begin{gather}
H(\mathbf{k})
=[\operatorname{Re}f_{0}(\mathbf{k})\,\rho_{x}
-\operatorname{Im}f_{0}(\mathbf{k})\,\rho_{y}]\sigma_{0}
+[\operatorname{Re}f_{z}(\mathbf{k})\,\rho_{x}
-\operatorname{Im}f_{z}(\mathbf{k})\,\rho_{y}]\sigma_{z}
+J_0\,\rho_{0}\sigma_{x}
\label{eq:H4band}
\\
= \begin{pmatrix}
0 & J_0 & f_0(\mathbf{k}) + f_z(\mathbf{k}) & 0 \\
J_0 & 0 & 0 & f_0(\mathbf{k}) - f_z(\mathbf{k}) \\
f_0^*(\mathbf{k}) + f_z^*(\mathbf{k}) & 0 & 0 & J_0 \\
0 & f_0^*(\mathbf{k}) - f_z^*(\mathbf{k}) & J_0 & 0
\end{pmatrix}
\end{gather}
with
$
f_{0}(\mathbf{k})
=-2\sqrt{2}\, t\cos (\tfrac{k_y}{2})\cos (\tfrac{k_x}{2})
e^{i\frac{k_x-k_y}{2}}=F_0(\mathbf{k})e^{i\phi_\mathbf{k}}
$,
$
f_{z}(\mathbf{k})
=-2\sqrt{2}\, t\cos (\tfrac{k_y}{2})\sin (\tfrac{k_x}{2})
e^{i\frac{k_x-k_y}{2}}=F_z(\mathbf{k})e^{i\phi_\mathbf{k}}.
$
The above Hamiltonian can be analytically diagonalized. We focus on one
of the sublattice branches. Furthermore, assuming
$|F_z(\mathbf{k})|\gg J_0$, we can effectively separate the spin-up and
down branches of the energy eigenvalues as,
\begin{equation}
\epsilon_{\uparrow(\downarrow)}(\mathbf{k})= F_0(\mathbf{k}) \pm
\big( F_z(\mathbf{k}) +\frac{J_0^2}{2F_z(\mathbf{k})}\big)
\end{equation}
We note that the assumption $|F_{z}(\mathbf{k})|\gg J_0$ effectively
decouples the two spin branches, which allows us to approximately
define the spin number along $z$. Recall that we have multiplied the
electron operators by the position-dependent phase for spin $\uparrow$
($\downarrow$), which acts as a plane wave and therefore shifts the
momentum label in a spin-dependent manner,
$c_{\mathbf{k},\sigma}=d_{\mathbf{k}+\sigma\frac{\pi}{2}\hat{x},\sigma}$.
The physical ($c$-electron) dispersion at momentum $\mathbf{k}$ is thus
read off from the $d$-electron bands at the shifted momentum,
 
\begin{equation}
\epsilon^c_{\uparrow(\downarrow)}(\mathbf{k})
=\epsilon_{\uparrow(\downarrow)}^d(k_x\pm \tfrac{\pi}{2},k_y)=
-4t \cos(\tfrac{k_x}{2})\cos(\tfrac{k_y}{2})
-\frac{J_0^{2}\cos\frac{k_{x}}{2}}{4t\cos\frac{k_{y}}{2}\cos k_{x}}
\pm\,\frac{J_0^{2}\sin\frac{k_{x}}{2}}{4t\cos\frac{k_{y}}{2}\cos k_{x}}
\end{equation}
By keeping the lowest harmonics, we derive the effective Hamiltonian of
the $p$-wave magnet as,
\begin{equation}
h_{\textrm{eff}}(\mathbf{k}) =
\big[\tfrac{t}{2}k^2-4t-\tfrac{J_0^2}{4t}\big]\sigma_0
+ \frac{J^2_{0}}{8t}\, k_x\, \sigma_z
\label{eqs:heff}
\end{equation}
This effective Hamiltonian possesses the effective time-reversal
symmetry $\Theta_\textrm{eff}=i\sigma_y\mathcal{K}$, which is the
consequence of $\mathcal{T'}$ in the lattice model. The half
translation $t_{\hat{x}}$ reduces to the spin-dependent momentum shift
on the low-energy bands and is absorbed by the gauge transformation
above. From now on, we take the effective two-band model of the
$p$-wave magnets to study the superconductivity.
Fig.~\ref{figs:lattice}(b),(c) show that the effective two-band model
quantitatively reproduces the band structure of the full lattice model
near the band bottom. This confirms the $p$-wave exchange coupling
derived above.

\section{Mean-field decomposition}
\label{sec:model}
\subsection{Symmetry classification and channel decoupling}
\label{sec:irrep_decoupling}
The point group of the p-wave magnet is $C_{2v}$. Each basis function in
Eq.~\eqref{eq:gap_decomposed} then transforms under a definite irreducible representation of
$C_{2v}$, as summarized in Table~\ref{tab:irrep_basis}. The singlet components ($\psi$,
$\Delta_x$, $\Delta_y$) and the triplet $\eta$ share the A$_1$ irrep and are
symmetry-allowed to mix, since the fourfold rotation that would separate the
$d_{x^2-y^2}$ combination $\Delta_x-\Delta_y$ from the $s$-wave sector is broken by the
exchange field, while the $p_y$-triplet $\Delta_{p_y}$ belongs to B$_2$ and decouples.
\begin{table}[h]
\caption{Classification of pairing under $C_{2v}$, acting jointly on the basis function and
the spin structure.}
\label{tab:irrep_basis}
\begin{ruledtabular}
\begin{tabular}{llll}
Component & Basis function $f_\alpha(\bk)$ & Parity ($\bk\to-\bk$) & $C_{2v}$ irrep \\
\hline
$\psi$  & $1$ & even & A$_1$ \\
$\Delta_x$  & $\cos k_x$ & even & A$_1$ \\
$\Delta_y$  & $\cos k_y$ & even & A$_1$ \\
$\eta$  & $\sin k_x$ & odd & A$_1$ \\
\hline
$\Delta_{p_y}$ & $\sin k_y$ & odd & B$_2$ \\
\end{tabular}
\end{ruledtabular}
\end{table}

Motivated by Eq.\eqref{eqs:heff}, we consider a tight-binding model on a two-dimensional square lattice.
The normal-state Hamiltonian is written as,
\begin{equation}
h_{\bk} = \xik\,\sigma_0 + \Jk\,\sigma_z,
\label{eq:hk}
\end{equation}
where $\xik = -2t(\cos k_x + \cos k_y) - \mu$. $\Jk = J\sin k_x$.
Here $t$ is the nearest-neighbor hopping amplitude. $\mu$ is the chemical potential. $J$ is the p-wave exchange term. The band dispersions of the normal Hamiltonian are $\epsilon_{\bk}^{\up/\dn} = \xik \pm \Jk$, satisfying the odd parity constraint $\epsilon_{\bk\up} = \epsilon_{-\bk\dn}$. We include the extended Hubbard interaction
\begin{equation}
H_I = -U\sum_i n_{i\uparrow}n_{i\downarrow} - V\sum_{\langle i,j\rangle,\sigma,\sigma'} n_{i\sigma}n_{j\sigma'},
\label{eq:HI_SM}
\end{equation}
where $n_{i\sigma} = c_{i\sigma}^\dagger c_{i\sigma}$ is the density operator for spin $\sigma$ at site $i$, with $U > 0$ and $V > 0$ representing the on-site and nearest-neighbor attractive strengths respectively. 

To analyze the pairing instabilities, we transform the interaction Hamiltonian into momentum space. In the BCS channel, the interaction is dominated by pairs with zero total momentum, which can be expressed as

\begin{equation}
H_I \approx -\frac{1}{N}\sum_{\mathbf{k},\mathbf{k}',\sigma,\sigma'} V_{\sigma\sigma'}(\mathbf{k} - \mathbf{k}') c_{\mathbf{k}\sigma}^\dagger c_{-\mathbf{k}\sigma'}^\dagger c_{-\mathbf{k}'\sigma'} c_{\mathbf{k}'\sigma}.
\label{eq:HI_kspaces}
\end{equation}
The interaction kernel $V_{\sigma\sigma'}(\mathbf{k} - \mathbf{k}')$ in Eq.~\eqref{eq:HI_kspaces} is obtained by the Fourier transform of the real-space coupling constants, yielding
\begin{equation}
V_{\sigma\sigma'}(\mathbf{k} - \mathbf{k}') = U(1-\delta_{\sigma\sigma'}) + 2V(\cos (k_x-k_x') + \cos (k_y-k_y')).
\label{eq:Vq}
\end{equation}

We now perform a mean-field decomposition of the interaction Hamiltonian. Focusing on the inter-spin pairing ($\sigma\neq\sigma'$), we apply the mean-field approximation $c_{\mathbf{k}\uparrow}^\dagger c_{-\mathbf{k}\downarrow}^\dagger c_{-\mathbf{k}'\downarrow} c_{\mathbf{k}'\uparrow} \approx \langle c_{\mathbf{k}\uparrow}^\dagger c_{-\mathbf{k}\downarrow}^\dagger \rangle c_{-\mathbf{k}'\downarrow} c_{\mathbf{k}'\uparrow} + c_{\mathbf{k}\uparrow}^\dagger c_{-\mathbf{k}\downarrow}^\dagger \langle c_{-\mathbf{k}'\downarrow} c_{\mathbf{k}'\uparrow} \rangle - \langle c_{\mathbf{k}\uparrow}^\dagger c_{-\mathbf{k}\downarrow}^\dagger \rangle \langle c_{-\mathbf{k}'\downarrow} c_{\mathbf{k}'\uparrow} \rangle$. The resulting mean-field Hamiltonian is
\begin{equation}
H_{\mathrm{MF}} = \sum_{\mathbf{k}} \left[ \Delta_{\uparrow\downarrow}(\mathbf{k})\,c_{\mathbf{k}\uparrow}^\dagger c_{-\mathbf{k}\downarrow}^\dagger + \mathrm{h.c.} \right] + \mathrm{const.},
\label{eq:HMF}
\end{equation}
where the superconducting gap function is defined as\begin{equation}
\Delta_{\uparrow\downarrow}(\mathbf{k}) = -\frac{1}{N}\sum_{\mathbf{k}'} V_{\uparrow\downarrow}(\mathbf{k} - \mathbf{k}') \langle c_{-\mathbf{k}'\downarrow} c_{\mathbf{k}'\uparrow} \rangle.
\label{eq:gap_def}
\end{equation}

\subsection{Pairing channel decomposition}
To make the channel decomposition explicit, we decompose the interaction kernel $V_{\uparrow\downarrow}(\mathbf{k} - \mathbf{k}')$, using the trigonometric identity $\cos(k_i - k_i') = \cos k_i \cos k_i' + \sin k_i \sin k_i'$. The interaction from Eq.~\eqref{eq:Vq} is expanded as,
\begin{equation}
V_{\uparrow\downarrow}(\mathbf{k} - \mathbf{k}') = U + 2V \sum_{i=x,y} \left( \cos k_i \cos k_i' + \sin k_i \sin k_i' \right),
\label{eq:V_expanded}
\end{equation}
which is separable in the five lattice harmonics $\{1,\,\cos k_x,\,\cos k_y,\,\sin k_x,\,\sin k_y\}$. Substituting Eq.~\eqref{eq:V_expanded} into the gap equation~\eqref{eq:gap_def}, the gap function decomposes as
\begin{equation}
\Delta_{\up\dn}(\bk) = \psi + \Delta_x\cos k_x + \Delta_y\cos k_y + \eta\,\sin k_x + \Delta_{p_y}\,\sin k_y,
\label{eq:gap_decomposed}
\end{equation}
with the self-consistency conditions for each component:
\begin{align}
\psi &= -\frac{U}{N}\sum_{\bk'} \langle c_{-\bk'\dn}\, c_{\bk'\up} \rangle, \label{eq:psi_def}\\
\eta &= -\frac{2V}{N}\sum_{\bk'} \sin k_x'\, \langle c_{-\bk'\dn}\, c_{\bk'\up} \rangle,
\quad
\Delta_{p_y} = -\frac{2V}{N}\sum_{\bk'} \sin k_y'\, \langle c_{-\bk'\dn}\, c_{\bk'\up} \rangle, \label{eq:eta_def}\\
\Delta_{x} &= -\frac{2V}{N}\sum_{\bk'} \cos k_x'\, \langle c_{-\bk'\dn}\, c_{\bk'\up} \rangle,
\quad
\Delta_{y} = -\frac{2V}{N}\sum_{\bk'} \cos k_y'\, \langle c_{-\bk'\dn}\, c_{\bk'\up} \rangle. \label{eq:d_def}
\end{align}
As shown in Sec.~\ref{sec:irrep_decoupling}, the four components $(\psi, \Delta_x, \Delta_y, \eta)$ belong to the same A$_1$ irrep and must be solved as a coupled problem, while the $p_y$-triplet $\Delta_{p_y}$ belongs to B$_2$ and decouples. The corresponding coupled gap equations are presented in Eqs.~\eqref{eq:eta_sc}.

\section{Derivation of the gap equations}
\label{sec:A1gap}

\subsection{$A_1$ channel}

In the Nambu basis $\Psi_{\bk} = (c_{\bk\up},\, c_{\bk\dn},\, c_{-\bk\up}^\dagger,\, c_{-\bk\dn}^\dagger)^T$, the BdG Hamiltonian for inter-spin pairing with gap function $\Delta_{\up\dn}(\bk)$ is
\begin{equation}
\mathcal{H}_{\mathrm{BdG}}(\bk) = \begin{pmatrix}
\xik+\Jk & 0 & 0 & \Delta_{\up\dn}(\bk) \\
0 & \xik-\Jk & \Delta_{\dn\up}(\bk) & 0 \\
0 & \Delta_{\dn\up}^*(\bk) & -\xik-J_{-\bk} & 0 \\
\Delta_{\up\dn}^*(\bk) & 0 & 0 & -\xik+J_{-\bk}
\end{pmatrix}.
\label{eq:BdG4x4}
\end{equation}

The $4\times4$ Hamiltonian block-diagonalizes into two independent $2\times2$ sectors. Using the p-wave property $J_{-\bk} = -\Jk$, the block Hamiltonians are given as,
\begin{equation}
H^{(1)}_{\bk} 
= (\xik + \Jk)\tau_z + \Delta_{\up\dn}(\bk)\,\tau_x,\quad\quad
H^{(2)}_{\bk} 
= (\xik - \Jk)\,\tau_z -\Delta_{\dn\up}(\bk)\tau_x,
\label{eq:block2}
\end{equation}
The corresponding eigenvalues of each block are given as,
\begin{align}
E^{(1)}_{\bk} = \pm\sqrt{(\xik + \Jk)^2 + |\Delta_{\up\dn}(\bk)|^2}, \quad
E^{(2)}_{\bk} = \pm\sqrt{(\xik - \Jk)^2 + |\Delta_{\dn\up}(\bk)|^2}. \label{eq:E2}
\end{align}
Each BdG energy band with the inter-spin pairing is fully gapped. We note that this is different from the case of the even parity altermagnet, which forms the Bogoliubov Fermi surface \cite{Hong2025}.

As discussed in Sec.~\ref{sec:irrep_decoupling}, the A$_1$ channel contains four symmetry-allowed components. The full A$_1$ gap function is therefore
\begin{equation}
\Delta_{\up\dn}(\bk) = \psi + \Delta_x\cos k_x + \Delta_y\cos k_y + \eta\sin k_x.
\label{eq:A1gap_full}
\end{equation}
By the anticommutation relation of fermions, $\Delta_{\dn\up}(\bk) = -\Delta_{\up\dn}(-\bk) = -(\psi + \Delta_x\cos k_x + \Delta_y\cos k_y) + \eta\sin k_x$.

We perform the standard Bogoliubov transformation for the first block.
Writing $H^{(1)}_{\bk} = (\xik + \Jk)\,\tau_z +  \Delta_{\up\dn}(\bk)\,\tau_x$ with real $\Delta_1(\bk) = \psi + \Delta_x\cos k_x + \Delta_y\cos k_y + \eta\sin k_x$, the transformation is
\begin{equation}
\begin{pmatrix} \gamma_{\bk} \\ \gamma_{-\bk}^\dagger \end{pmatrix}
= \begin{pmatrix} u_{\bk} & -v_{\bk} \\ v_{\bk} & u_{\bk} \end{pmatrix}
\begin{pmatrix} c_{\bk\up} \\ c_{-\bk\dn}^\dagger \end{pmatrix},
\end{equation}
where $u_{\bk}^2 = \frac{1}{2}\left(1 + {\varepsilon_1}/{E^{(1)}_{\bk}}\right)$, 
$v_{\bk}^2 = \frac{1}{2}\left(1 - {\varepsilon_1}/{E^{(1)}_{\bk}}\right)$, 
$ u_{\bk} v_{\bk} = {\Delta_1}/{2E_{\bk}}$. The correlation function is given as,
\begin{equation}
\langle c_{-\bk\dn}\, c_{\bk\up} \rangle
= -u_{\bk} v_{\bk}\,(1 - 2f(E^{(1)}_{\bk}))
= -\frac{\Delta_1(\bk)}{2E^{(1)}_{\bk}}\,\tanh\frac{E^{(1)}_{\bk}}{2T},
\label{eq:anomalous}
\end{equation}
where $f(\epsilon) = 1/(1+e^{\epsilon/T})$ is the Fermi-Dirac distribution function.

We derive the self-consistent gap equations. From Eqs.~\eqref{eq:psi_def}, \eqref{eq:eta_def}, and~\eqref{eq:d_def}, we obtain the following relations,
\begin{align}
\psi &= \frac{U}{N}\sum_{\bk} \frac{\Delta_1(\bk)}{2E^{(1)}_{\bk}}\,\tanh\frac{E^{(1)}_{\bk}}{2T},  \quad
\Delta_{i} = \frac{2V}{N}\sum_{\bk} \frac{\cos k_i\,\Delta_1(\bk)}{2E^{(1)}_{\bk}}\,\tanh\frac{E^{(1)}_{\bk}}{2T}~~(i=x,y), \quad
\eta = \frac{2V}{N}\sum_{\bk} \frac{\sin k_x\,\Delta_1(\bk)}{2E^{(1)}_{\bk}}\,\tanh\frac{E^{(1)}_{\bk}}{2T}. \label{eq:eta_sc}
\end{align}

The above self-consistent equations can be rewritten in the compact $4\times4$ matrix form as,
\begin{equation}
\begin{pmatrix} \psi \\ \Delta_{x} \\ \Delta_{y} \\ \eta \end{pmatrix}
= \begin{pmatrix}
U\mathcal{A} & U\mathcal{X}_x & U\mathcal{X}_y & U\mathcal{B} \\
2V\mathcal{X}_x & 2V\mathcal{D}_{xx} & 2V\mathcal{D}_{xy} & 2V\mathcal{P}_x \\
2V\mathcal{X}_y & 2V\mathcal{D}_{xy} & 2V\mathcal{D}_{yy} & 2V\mathcal{P}_y \\
2V\mathcal{B} & 2V\mathcal{P}_x & 2V\mathcal{P}_y & 2V\mathcal{C}
\end{pmatrix}
\begin{pmatrix} \psi \\ \Delta_{x} \\ \Delta_{y} \\ \eta \end{pmatrix},
\label{eq:A1matrix}
\end{equation}
where the matrix elements are defined by the overlap integrals
\begin{align}
\mathcal{A}(T) &\equiv \frac{1}{N}\sum_{\bk} \frac{1}{2E^{(1)}_{\bk}}\,\tanh\frac{E^{(1)}_{\bk}}{2T}, \qquad
\mathcal{B}(T) \equiv \frac{1}{N}\sum_{\bk} \frac{\sin k_x}{2E^{(1)}_{\bk}}\,\tanh\frac{E^{(1)}_{\bk}}{2T}, \qquad
\mathcal{C}(T) \equiv \frac{1}{N}\sum_{\bk} \frac{\sin^2 k_x}{2E^{(1)}_{\bk}}\,\tanh\frac{E^{(1)}_{\bk}}{2T}, \nonumber\\
\mathcal{X}_i(T) &\equiv \frac{1}{N}\sum_{\bk} \frac{\cos k_i}{2E^{(1)}_{\bk}}\,\tanh\frac{E^{(1)}_{\bk}}{2T}, \qquad
\mathcal{D}_{ij}(T) \equiv \frac{1}{N}\sum_{\bk} \frac{\cos k_i \cos k_j}{2E^{(1)}_{\bk}}\,\tanh\frac{E^{(1)}_{\bk}}{2T}, \qquad
\mathcal{P}_i(T) \equiv \frac{1}{N}\sum_{\bk} \frac{\cos k_i \sin k_x}{2E^{(1)}_{\bk}}\,\tanh\frac{E^{(1)}_{\bk}}{2T}. \label{eq:calC}
\end{align}
The off-diagonal elements $\mathcal{B}$ and $\mathcal{P}_i$ couple the singlet and triplet components. At $J = 0$ the integrals additionally satisfy $\mathcal{X}_x = \mathcal{X}_y$ and $\mathcal{D}_{xx} = \mathcal{D}_{yy}$ by the fourfold rotation, so that the $d_{x^2-y^2}$ decouples from the $s$-wave sector.

\textit{Linearized gap equations}--
Near $T_c$, the gap amplitudes $\psi, \Delta_x, \Delta_y, \eta \to 0$ and the BdG quasiparticle energy reduces to the normal-state dispersion:
\begin{equation}
E^{(1)}_{\bk} = \sqrt{(\xik + \Jk)^2 + \Delta_1^2(\bk)}
\;\xrightarrow{\psi,\Delta_x,\Delta_y,\eta\to 0}\;
|\xik + \Jk|.
\end{equation}
The linearized gap equation for the full $4\times4$ A$_1$ system is the eigenvalue problem
\begin{equation}
M_0^{(4)}(T_c)\,\begin{pmatrix}\psi \\ \Delta_{x} \\ \Delta_{y} \\ \eta\end{pmatrix} = \begin{pmatrix}\psi \\ \Delta_{x} \\ \Delta_{y} \\ \eta\end{pmatrix},
\quad
M_0^{(4)} = \begin{pmatrix}
U\mathcal{A}_0 & U\mathcal{X}_{x,0} & U\mathcal{X}_{y,0} & U\mathcal{B}_0 \\
2V\mathcal{X}_{x,0} & 2V\mathcal{D}_{xx,0} & 2V\mathcal{D}_{xy,0} & 2V\mathcal{P}_{x,0} \\
2V\mathcal{X}_{y,0} & 2V\mathcal{D}_{xy,0} & 2V\mathcal{D}_{yy,0} & 2V\mathcal{P}_{y,0} \\
2V\mathcal{B}_0 & 2V\mathcal{P}_{x,0} & 2V\mathcal{P}_{y,0} & 2V\mathcal{C}_0
\end{pmatrix},
\label{eq:M0_4x4}
\end{equation}
where the subscript $0$ denotes the integrals in Eq.~\eqref{eq:calC} evaluated with $E^{(1)}_{\bk}\to|\xik+\Jk|$, and the critical temperature $T_c$ is determined by the condition that the largest eigenvalue of $M_0^{(4)}(T)$ equals unity,
\begin{equation}
\lambda_{\max}(T_c) = 1.
\label{eq:Tc_condition}
\end{equation}

To expose the singlet--triplet locking mechanism analytically, it is instructive to project onto the $(\psi,\eta)$ subspace, which captures the essential parity mixing:
\begin{equation}
M_0(T_c)\,\begin{pmatrix}\psi \\ \eta\end{pmatrix} = \begin{pmatrix}\psi \\ \eta\end{pmatrix},
\quad
M_0 = \begin{pmatrix} U\mathcal{A}_0 & U\mathcal{B}_0 \\ 2V\mathcal{B}_0 & 2V\mathcal{C}_0 \end{pmatrix}.
\label{eq:M0}
\end{equation}
The eigenvalues are given as
\begin{equation}
\lambda_\pm = \frac{1}{2}\left(U\mathcal{A}_0 + 2V\mathcal{C}_0\right) \pm \frac{1}{2}\sqrt{(U\mathcal{A}_0 - 2V\mathcal{C}_0)^2 + 8UV\mathcal{B}_0^2}.
\label{eq:lam_plus}
\end{equation}
The cross-term $8UV\mathcal{B}_0^2$ always increases $\lambda_+$, confirming that the singlet--triplet coupling enhances $T_c$.
Furthermore, the eigenvector corresponding to $\lambda_+$ gives the gap ratio
\begin{equation}
\frac{\eta}{\psi} = \frac{\lambda_+ - U\mathcal{A}_0}{U\mathcal{B}_0} = \frac{2V\mathcal{B}_0}{\lambda_+ - 2V\mathcal{C}_0}.
\label{eq:ratio_SM}
\end{equation}
When $U \gg V$: $\lambda_+ \approx U\mathcal{A}_0$ and $|\eta/\psi| \ll 1$ (singlet-dominant).
When $V \gg U$: $\lambda_+ \approx 2V\mathcal{C}_0$ and $|\eta/\psi| \gg 1$ (triplet-dominant).
The bond singlets enter this projection only through quantitative renormalizations of the singlet sector and do not modify the locking structure.

\subsection{$B_2$ channel ($p_y$-wave pairing) }

With nearest-neighbor interactions only, the B$_2$ channel supports a pure $p_y$-triplet state $\Delta_{\up\dn}(\bk) = \Delta_{p_y} \sin k_y$.
The first block Hamiltonian gives $\epsilon^{(1)}_{\bk} = \sqrt{(\xik + \Jk)^2 + \Delta_{p_y}^2\sin^2 k_y}$.
The linearized $T_c$ equation is
\begin{equation}
1 = 2V\,\frac{1}{N}\sum_{\bk} \frac{\sin^2 k_y\;\tanh\frac{|\xik + \Jk|}{2T_c}}{2|\xik + \Jk|}.
\label{eq:B2_Tc}
\end{equation}

\section{Gap equations in an external Zeeman field}
\label{sec:gap_zeeman}

\subsection{BdG Hamiltonian with Zeeman field}

The normal Hamiltonian with a general in-plane ($h_x$) and out-of-plane ($h_z$) Zeeman field is
\begin{equation}
h(\bk) = \xik\,\sigma_0 + (\Jk + h_z)\sigma_z + h_x\,\sigma_x\ .
\label{eq:hk_zeeman}
\end{equation}
The corresponding eigenenergies at $\bk$ are
$\varepsilon_\pm(\bk) = \xik \pm \sqrt{d_{\bk}^2 + h_x^2}$, where $d_{\bk}=\Jk + h_z$. The BdG Hamiltonian with the Zeeman field is
\begin{equation}
\mathcal{H}_{\mathrm{BdG}}(\bk)=H_0(\bk)+H_{pair}(\bk) = \begin{pmatrix}
h(\bk) & D(\bk) \\ D^\dagger(\bk) & -h^*(-\bk)
\end{pmatrix},
\label{eq:BdG_zeeman}
\end{equation}
where $D(\bk) = \bigl(\begin{smallmatrix} 0 & \Delta_{\up\dn} \\ \Delta_{\dn\up} & 0 \end{smallmatrix}\bigr)$ is the pairing matrix [cf.\ Eq.~\eqref{eq:BdG4x4}]. In the weak pairing limit, we define the Matsubara Green function for the BdG Hamiltonian as,
\begin{equation}
G(i\omega_n)=\frac{1}{i\omega_n-(H_0(\bk)+H_{pair}(\bk))}\approx G_0(i\omega_n) + G_0(i\omega_n)H_{pair}(\bk) G_0(i\omega_n)
\end{equation}
where $G_0(i\omega_n)=\frac{1}{i\omega_n-H_0(\bk)}=\begin{pmatrix}
g_{e}(\mathbf{k}, i\omega_n) & 0 \\ 0 & g_{h}(\mathbf{k}, i\omega_n)
\end{pmatrix}$ is the normal state Green function. 

We compute the anomalous term $\langle c_{-\bk\dn}\,c_{\bk\up}\rangle$ in the mean-field state. 
\begin{alignat}{2}
\langle c_{-\mathbf{k} \downarrow} c_{\mathbf{k} \uparrow} \rangle =& T \sum_{n} \left[ g_{e, \uparrow\uparrow}(\mathbf{k}, i\omega_n) \Delta_{\uparrow\downarrow}(\mathbf{k}) g_{h, \downarrow\downarrow}(\mathbf{k}, i\omega_n) + g_{e, \uparrow\downarrow}(\mathbf{k}, i\omega_n) \Delta_{\downarrow\uparrow}(\mathbf{k}) g_{h, \uparrow\downarrow}(\mathbf{k}, i\omega_n) \right]
\\
 =& T \sum_n \Big\{ \Delta_s(\mathbf{k}) \big[ g_{e, \uparrow\uparrow} g_{h, \downarrow\downarrow} - g_{e, \uparrow\downarrow} g_{h, \uparrow\downarrow} \big] + \eta \sin k_x \big[ g_{e, \uparrow\uparrow} g_{h, \downarrow\downarrow} + g_{e, \uparrow\downarrow} g_{h, \uparrow\downarrow} \big] \Big\}
\end{alignat}
We have inserted the A$_1$ gap function $\Delta_{\up\dn}(\bk) = \Delta_s(\bk) + \eta\sin k_x$ with the singlet part $\Delta_s(\bk) \equiv \psi + \Delta_x\cos k_x + \Delta_y\cos k_y$, and its Pauli partner $\Delta_{\dn\up}(\bk) = -\Delta_s(\bk) + \eta\sin k_x$. Using the definitions~\eqref{eq:psi_def}--\eqref{eq:d_def}, the self-consistent gap equations become
\begin{align}
\psi =& 
-\frac{UT}{N} \sum_{\mathbf{k}, n}  \left\{ \Delta_s(\mathbf{k}) \big[ g_{e, \uparrow\uparrow} g_{h, \downarrow\downarrow} - g_{e, \uparrow\downarrow} g_{h, \uparrow\downarrow} \big] + \eta \sin k_x \big[ g_{e, \uparrow\uparrow} g_{h, \downarrow\downarrow} + g_{e, \uparrow\downarrow} g_{h, \uparrow\downarrow} \big] \right\},\label{eq:psi_sc_h}\\
\Delta_{i} =& 
-\frac{2VT}{N} \sum_{\mathbf{k}, n}  \cos k_i \left\{ \Delta_s(\mathbf{k}) \big[ g_{e, \uparrow\uparrow} g_{h, \downarrow\downarrow} - g_{e, \uparrow\downarrow} g_{h, \uparrow\downarrow} \big] + \eta \sin k_x \big[ g_{e, \uparrow\uparrow} g_{h, \downarrow\downarrow} + g_{e, \uparrow\downarrow} g_{h, \uparrow\downarrow} \big] \right\}~~(i=x,y), \label{eq:es_sc_h}\\
\eta = &
-\frac{2VT}{N} \sum_{\mathbf{k}, n}  \sin k_x \left\{ \Delta_s(\mathbf{k}) \big[ g_{e, \uparrow\uparrow} g_{h, \downarrow\downarrow} - g_{e, \uparrow\downarrow} g_{h, \uparrow\downarrow} \big] + \eta \sin k_x \big[ g_{e, \uparrow\uparrow} g_{h, \downarrow\downarrow} + g_{e, \uparrow\downarrow} g_{h, \uparrow\downarrow} \big] \right\}. \label{eq:eta_sc_h}
\end{align}
Near $T_c$, the gap amplitudes vanish ($\psi, \Delta_x, \Delta_y, \eta \to 0$) and $g_e$, $g_h$ are evaluated with the normal-state dispersions.
The self-consistent equations~\eqref{eq:psi_sc_h}--\eqref{eq:eta_sc_h} linearize into the $4\times 4$ eigenvalue problem again [cf.\ Eq.~\eqref{eq:M0_4x4}].

For the $B_2$ channel, we derive the similar result as,
\begin{equation}
1 = -\frac{2VT}{N}\sum_{\bk,n} \sin^2 k_y\;\big[g_{e,\up\up}\,g_{h,\dn\dn} + g_{e,\up\dn}\,g_{h,\up\dn}\big]\,.
\label{eq:B2_Tc_h}
\end{equation}

\section{Topological classification}
\label{sec:Z2}

\subsection{First order topological invariant in Class DIII}

The full A$_1$ BdG Hamiltonian corresponds to class DIII in Altland--Zirnbauer classification, which supports a $\mathbb{Z}_2$ topological invariant. However, we find that the A$_1$ Ising s$+$p state has $\mathbb{Z}_2 = 0$. The $\mathbb{Z}_2$ invariant is determined by the sign of the Pfaffian of $w(\bk) = \langle u_{-\bk} | \mathcal{T} | u_{\bk} \rangle$ at the four time-reversal invariant momenta (TRIM): $\Gamma = (0,0)$, $X = (\pi,0)$, $Y = (0,\pi)$, $M = (\pi,\pi)$.
At all TRIM, $\sin k_x = 0$, so $\Jk = J\sin k_x = 0$ and $\Delta_1 = \psi + \eta\sin k_x = \psi > 0$.
The gap does not change sign between any pair of TRIM, yielding
\begin{equation}
(-1)^{\nu} = \prod_{\text{TRIM}} \sgn[\Delta(\bk_i)] = (+1)^4 = +1 
\end{equation}

\subsection{ 1D winding number}
\label{sec:winding}

Each block of the A$_1$ BdG Hamiltonian has the form $H = \varepsilon\,\tau_z + \Delta\,\tau_x$.
This anticommutes with $\mathcal{S} = \tau_y$:
\begin{equation}
\{\mathcal{S},\, H^{(1)}_{\bk}\} = 0, \quad \mathcal{S} = \tau_y.
\end{equation}
This chiral symmetry endows the system with a 1D $\mathbb{Z}$ topological invariant (winding number) at each fixed $k_y$.

In the eigenbasis of $\mathcal{S} = \tau_y$, the Hamiltonian becomes purely off-diagonal, allowing to define the complex function
\begin{equation}
q(k_x; k_y) \equiv \varepsilon_1(k_x, k_y) + i\,\Delta_1(k_x),
\label{eq:q_def}
\end{equation}
where
\begin{align}
\varepsilon_1(k_x, k_y) = -2t(\cos k_x + \cos k_y) - \mu + J\sin k_x, \quad
\Delta_1(k_x) &= \psi + \eta\sin k_x.
\end{align}
The 1D winding number at fixed $k_y$ is defined as
\begin{equation}
\nu(k_y) = \frac{1}{2\pi i}\oint_{-\pi}^{\pi} dk_x\,\frac{1}{q}\frac{dq}{dk_x}
= \frac{1}{2\pi}\oint d(\arg q),
\label{eq:winding_def}
\end{equation}
which counts the number of times $q(k_x)$ winds around the origin in the complex plane as $k_x$ traverses the Brillouin zone. For $\nu(k_y) \neq 0$, the curve $q(k_x)$ must encircle the origin.

At the gap zeros, $\sin k_x^* = -\psi/\eta$, so
\begin{equation}
\cos k_x^{(1)} = +\alpha, \quad \cos k_x^{(2)} = -\alpha, \quad \alpha \equiv \sqrt{1 - (\psi/\eta)^2}.
\end{equation}
The dispersions at the gap zeros are
\begin{align}
\varepsilon_1(k_x^{(1)}, k_y) &= -2t\alpha - 2t\cos k_y - \mu - J\psi/\eta \equiv -2t\alpha - \beta(k_y), \\
\varepsilon_1(k_x^{(2)}, k_y) &= +2t\alpha - 2t\cos k_y - \mu - J\psi/\eta \equiv +2t\alpha - \beta(k_y),
\end{align}
where we define
\begin{equation}
\beta(k_y) \equiv 2t\cos k_y + \mu + J\psi/\eta.
\end{equation}

The condition for opposite signs, $\varepsilon_1^{(1)} \cdot \varepsilon_1^{(2)} < 0$, becomes $(2t\alpha)^2 > \beta(k_y)^2$, i.e.,
\begin{equation}
\nu(k_y) \neq 0 \;\Longleftrightarrow\; |\beta(k_y)| < 2t\alpha = 2t\sqrt{1 - (\psi/\eta)^2}.
\label{eq:winding_criterion}
\end{equation}

At the mirror-invariant momenta $k_y = 0$ and $k_y = \pi$:
\begin{align}
k_y = 0:& \quad |\mu + J\psi/\eta + 2t| < 2t\sqrt{1 - (\psi/\eta)^2}, \label{eq:nu0} \\
k_y = \pi:& \quad |\mu + J\psi/\eta - 2t| < 2t\sqrt{1 - (\psi/\eta)^2}. \label{eq:nupi}
\end{align}

Defining $r \equiv \psi/\eta$ (with $|r| < 1$), $\alpha = \sqrt{1 - r^2}$, and $\beta_0 \equiv \mu + Jr$:

\textit{Region I} ($\nu(0) = 1$, $\nu(\pi) = 0$):
\begin{equation}
-2t(1 + \alpha) < \beta_0 < -2t(1 - \alpha) \quad \text{and} \quad |\beta_0 - 2t| \geq 2t\alpha.
\end{equation}

\textit{Region II} ($\nu(0) = 0$, $\nu(\pi) = 1$):
\begin{equation}
2t(1 - \alpha) < \beta_0 < 2t(1 + \alpha) \quad \text{and} \quad |\beta_0 + 2t| \geq 2t\alpha.
\end{equation}

\textit{Trivial} ($\nu(0) = \nu(\pi)$):
the complement of Regions I and II.

Block 2 [Eq.~\eqref{eq:block2}] has gap function $\Delta_2(k_x) = \psi - \eta\sin k_x$, with zeros at $\sin k_x = +\psi/\eta$.
The corresponding dispersions at the gap zeros are
\begin{align}
\varepsilon_2(k_x^{(1)}, k_y) &= -2t\alpha - \beta(k_y), \\
\varepsilon_2(k_x^{(2)}, k_y) &= +2t\alpha - \beta(k_y),
\end{align}
which are identical to the Block 1 expressions.
Therefore $|\nu_1(k_y)| = |\nu_2(k_y)|$ at all $k_y$.

Each BdG block has quasiparticle energy $E^{(1)}_{\bk} = \sqrt{(\varepsilon_1)^2 + \Delta_1^2}$, where $\varepsilon_1 = \xik + \Jk$ and $\Delta_1 = \psi + \eta\sin k_x$.
This energy vanishes when both $\varepsilon_1 = 0$ (Fermi surface) and $\Delta_1 = 0$ (gap zero at $\sin k_x^* = -\psi/\eta$) are simultaneously satisfied. The node locations coincide with the momenta where the one-dimensional winding number $\nu(k_y)$ undergoes a topological transition from $0$ to $1$. Since the quasiparticle density of states vanishes linearly near the nodes, $\rho(E) \propto |E|$, analogous to a $d$-wave superconductor. However, the bulk gap is nonzero at the time-reversal--invariant momenta ($k_y = 0, \pi$) where the winding number is evaluated, ensuring that $\nu(0)$ and $\nu(\pi)$ are individually well-defined.

\newpage 
\pagebreak

\begin{figure*}[t]
\centering
\includegraphics[width=1\linewidth]{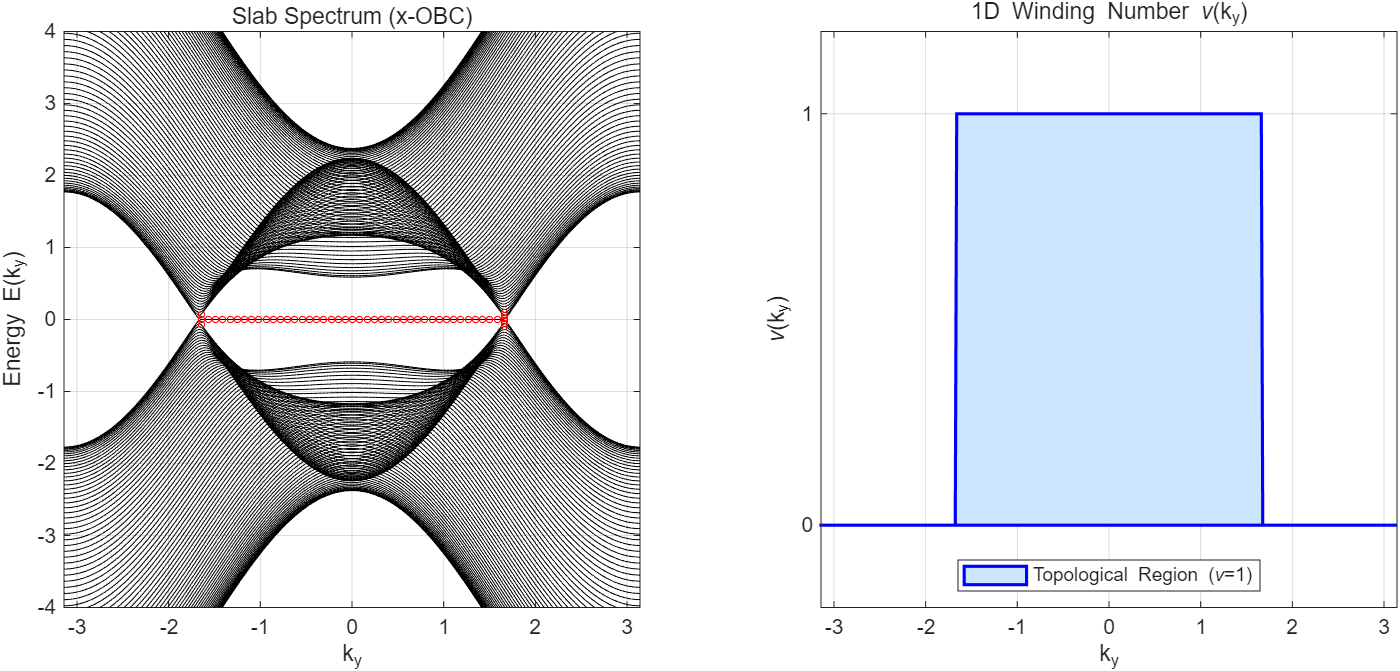}
\caption{
Topological properties of the $A_1$ channel superconducting state. (left)  Energy spectrum $E(k_y)$ calculated under open boundary conditions (OBC) in the $x$-direction and periodic boundary conditions in the $y$-direction. The red lines localized at $E=0$ represent the topologically protected Majorana edge modes. (right) The 1D winding number $\nu(k_y)$ calculated as a function of $k_y$. The topological region matches the momentum range in where the zero-energy edge states appear.
}
\label{fig:edge}
\end{figure*}

\begin{figure*}[t]
\centering
\includegraphics[width=1\linewidth]{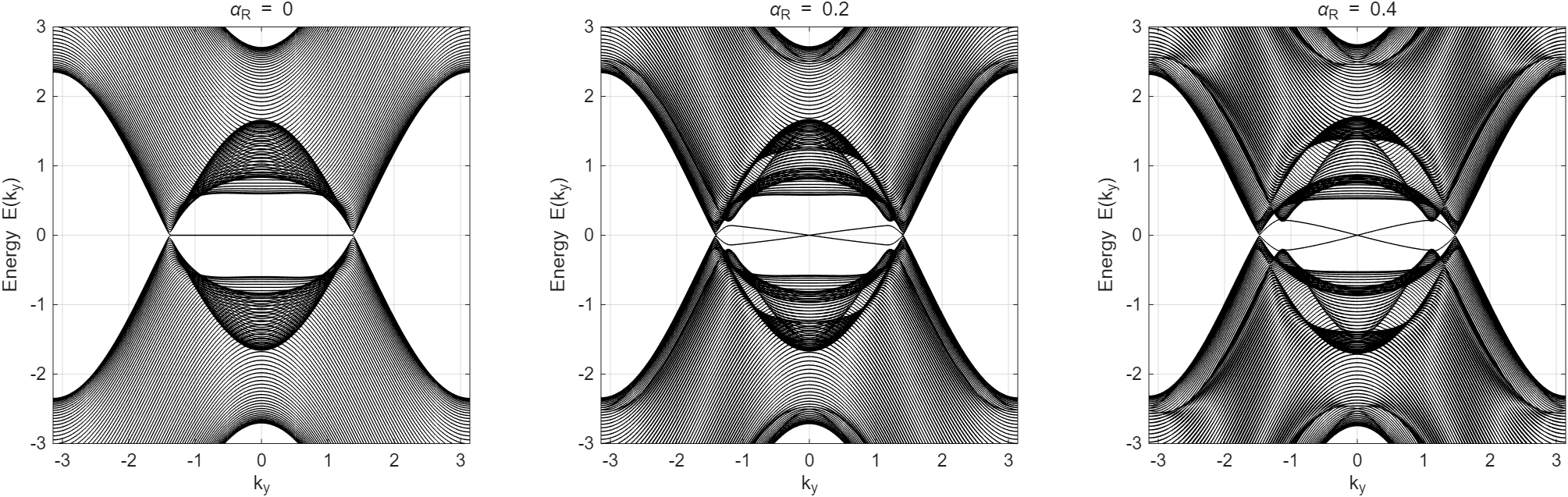}
\caption{
Evolution of the slab BdG spectrum $E(k_y)$ as a function of the effective spin-orbit coupling strength $\alpha_R$, calculated 
under open boundary conditions (OBC) in the $x$-direction and periodic 
boundary conditions in the $y$-direction. From left to right, 
$\alpha_R = 0$, $0.2$, and $0.4$. At $\alpha_R = 0$, the zero-energy 
modes form a flat band over a finite $k_y$ window, characteristic of 
the nodal class-DIII phase with momentum-resolved 1D winding number.
}
\label{fig:edge_soc}
\end{figure*}

\end{document}